\documentclass[aps,pre,reprint,doi=false,isbn=false,url=false,title=true,longbibliography,noeprint,footinbib]{revtex4-1}
\usepackage{graphicx}
\usepackage{amsmath}
\usepackage{array}
\usepackage{amsfonts}
\usepackage{float}
\usepackage{xcolor}
\usepackage[mathscr]{euscript}
\def\ba{\begin{eqnarray}}
\def\ea{\end{eqnarray}}
\def\be{\begin{equation}}
\def\ee{\end{equation}}
\def\bm{\begin{math}}
\def\me{\end{math}}

\newcommand{\dummy}

\makeatletter
\newcommand{\fmarki}{*}
\newcommand{\fmarkii}{\ensuremath{\dagger}}
\newcommand{\fmarkiii}{\ensuremath{\ddagger}}
\newcommand{\fmarkiv}{\ensuremath{\mathsection}}
\newcommand{\fmarkv}{\ensuremath{\mathparagraph}}
\newcommand{\fmarkvi}{\ensuremath{\|}}
\newcommand{\fmarkvii}{**}
\newcommand{\fmarkviii}{\ensuremath{\dagger\dagger}}
\newcommand{\fmarkix}{\ensuremath{\ddagger\ddagger}}
                
\def\@fnsymbol#1{{\ifcase#1\or \fmarki\or \fmarkii\or \fmarkiii\or \fmarkiv\or \fmarkv\or \fmarkvi\or \fmarkvii\or \fmarkviii\or \fmarkix \else\@ctrerr\fi}}
\makeatother

\renewcommand{\fmarki}{$\dagger$}
 \renewcommand{\fmarkii}{*}
 \renewcommand{\fmarkiii}{$\ddagger$}
 \renewcommand{\fmarkiv}{a$_4$}
 \renewcommand{\fmarkv}{x$_5$}

\begin{document}

\title{Interplay of phase segregation and chemical reaction: Crossover and effect on growth laws}
\author{Shubham Thwal}\email[]{shubhamthwal1@gmail.com}
\affiliation{Amity Institute of Applied Sciences, Amity University Uttar Pradesh, Noida 201313,
India
}
\author{Suman Majumder}\email[]{smajumder.@amity.edu, suman.jdv@gmail.com}

\affiliation{Amity Institute of Applied Sciences, Amity University Uttar Pradesh, Noida 201313,
India
}
\begin{abstract}
By combining the nonconserved spin-flip dynamics driving ferromagnetic ordering with the conserved Kawasaki-exchange dynamics driving phase segregation, we perform Monte Carlo simulations of the nearest neighbor Ising model. Such a set up mimics a system consisting of a binary mixture of \emph{isomers} which is simultaneously undergoing a segregation and an \emph{interconversion} reaction among themselves . Here, we study such a system following a quench from the high-temperature homogeneous phase to a temperature below the demixing transition. We monitor the growth of domains of both the \emph{winner}, the \emph{isomer} which survives as the majority and the \emph{loser}, the \emph{isomer} that perishes. Our results show a strong interplay of the two dynamics at early times leading to a growth of the average domain size of both the \emph{winner} and \emph{loser} as $\sim t^{1/7}$, slower than a purely phase-segregating system. At later times, eventually the dynamics becomes reaction dominated, and the \emph{winner} exhibits a  $\sim t^{1/2}$ growth, expected for a system with purely nonconserved dynamics. On the other hand, the \emph{loser} at first show a faster growth, albeit, slower than the \emph{winner}, and then starts to decay before it almost vanishes. Further, we estimate the time $\tau_s$  marking the crossover from the early-time slow growth to the late-time reaction dominated faster growth. As a function of the reaction probability $p_r$, we observe a power-law scaling $\tau_s \sim p_r^{-x}$, where $x\approx 1.05$, irrespective of temperature. For a fixed value of $p_r$ too, $\tau_s$ appears to be independent of temperature.
\end{abstract}

\maketitle

\section{INTRODUCTION}
Nonequilibrium dynamics of a system following a quench from a high-temperature disordered state to a temperature below the critical temperature $T_c$, marking an order-disorder phase transition, has been an active research topic over the last five decades 
\cite{bray2002,puri_book}. Typical examples of order-disorder transitions are ferromagnetic ordering and phase segregation in a binary mixture. The equilibrium aspects of these transitions, viz., values of static critical exponents, bear universal features \cite{stanley1971,fisher1974,Hohenberg1977,domb2000}. On the other hand, the corresponding nonequilibrium kinetics of the transitions may belong to different universality classes depending on the intrinsic transport mechanism of the system \cite{Siggia1979,Furukawa1985,Furukawa1987}. However, phenomenology  of both the transitions is highlighted by the formation and growth of domains of aligned magnetic spins or like-species.  Importantly, the concerned domain growth  is a scaling phenomenon \cite{bray2002,puri_book}, i.e., various morphology-characterizing functions, viz., the two-point equal-time order parameter correlation function $C(r,t)$ and the structure factor $S(k,t)$, respectively, obey the relations
\begin{equation}\label{cor_scaling}
 C(r,t) \equiv \tilde{C}(x);~ x=r/\ell(t),
\end{equation}
and
\begin{equation}\label{skt_scaling}
 S(k,t) \equiv \ell(t)^d\tilde{S}(x);~x=k\ell(t),
\end{equation}
where $\tilde{C}(x)$ and $\tilde{S}(x)$ are time ($t$)-independent scaling functions, $d$ is the system dimensionality, and $\ell(t)$ is the time-dependent characteristic length scale measuring the average domain size.   In general, $\ell(t)$ grows in a power-law fashion as 
\begin{eqnarray}\label{power-law}
\ell(t) \sim t^{\alpha}.
\end{eqnarray}
The growth exponent $\alpha$ depends on the intrinsic dynamics of the system. While for ferromagnetic ordering $\alpha=1/2$ representing the Lifshitz-Cahn-Allen (LCA) law \cite{allen1979}, for a solid binary-mixture phase segregation $\alpha=1/3$ stands for the Lifshitz-Slyozov (LS) law \cite{lifshitz1961,lifshitz1962}. 
\par
In ferromagnetic ordering or phase ordering, typically one starts with a system having magnetization $m\approx0$, and after the quench the system approaches its new equilibrium state where $m \neq 0$. The intrinsic dynamics of this phenomenon is nonconserved, as the order parameter $m$ changes continuously during the evolution. Conversely, during phase segregation in a binary mixture following a quench from a homogeneous or miscible phase, the order parameter, i.e., the concentration difference $\chi$ between the two species, remains constant throughout the evolution. Thus, its dynamics is said to be conserved. Monte Carlo simulations (MC) of the nearest neighbor Ising model has been used extensively to study domain growth in both conserved and nonconserved systems, and the corresponding 
growth laws have been verified \cite{bray2002,marko1995,amar1988,majumder2010,majumder2011,majumder2013}. Recently, the interest has shifted toward investigating domain growth in more complex and computationally demanding systems, e.g.,the long-range Ising model with power-law interaction, using both conserved and nonconserved order-parameter dynamics \cite{christiansen2019,christiansen2020,christiansen2021,mueller2022}.
\par
In this work, by means of  MC simulations of the Ising model, we investigate the effects of combining conserved and nonconserved dynamics, on the domain-growth laws. The motivation of studying such a system stems from understanding the nonequilbrium dynamics of a binary mixture where an \emph{interconversion} reaction among the components occurs simultaneously with segregation among themselves. 
Typical example of such a system undergoing \emph{interconversion}  reaction is  \emph{isomeric} mixtures, e.g., conversion of \emph{cis-isomer} to \emph{trans-isomer}, or conversion between optical \emph {isomers} or \emph{enantiomers} \cite{mcnaught_book}. 
Either naturally or due to some external drive the components of the \emph{isomeric} mixture do segregate from each other, and in combination with the \emph{interconversion} reaction the mixture gets enriched with one of the \emph{isomers} \cite{soai1995,shibata1996,shibata1996_jacs,tran1996reaction,Qui1997,ohta1998,viedma2005,lombardo2009}. There have been few studies on this kind of systems either by using MC simulations of the Ising model or by solving the corresponding Cahn-Hilliard-Cook (CHC) equation \cite{glotzer1994,puri1994,glotzer1995,singh2012,shumovskyi2021}. While most of the studies were motivated to understand the amplification of one of the phases, the CHC approach focused on the domain-growth laws reporting a crossover from early-time reaction controlled LCA behavior to a late-time diffusion controlled LS behavior \cite{singh2012}. Interest has also been laid on considering reactions in solution via molecular dynamics (MD) simulations \cite{longo2022}. Even studies with forceful preservation of the compositions being imposed along with the \emph{interconversion} reaction have also been conducted, leading to fascinating pattern formation mimicking morphology of microphase separation found in the biological world \cite{longo2023}. Lately, using the Ising model it has been shown that the Arrhenius behavior of the \emph{interconversion}  gets significantly affected due to the simultaneous existence of the segregation process \cite{Thwal1}. 
\par
The effective dynamics of the system described above is nonconserved, i.e., in the final state one of the reacting species will survive as the majority which we refer to as the \emph{winner} and the other one as \emph{loser}. Recently, one of us has demonstrated how to disentangle the growth of the \emph{winner} from the \emph{loser} during phase ordering \cite{majumder2023}. Importantly, such a disentanglement allows one to realize the scaling laws associated with domain growth using systems with relatively smaller size. Here, adopting the same disentanglement protocol we study the growth of the \emph{winner} and \emph{loser} in a binary mixture of \emph{isomers} which is undergoing both segregation and an \emph{interconversion} reaction. This not only allows us to have an appropriate realization of the associated domain-growth laws, but also to extract a timescale that marks the crossover between the two types of dynamics present in the system.  
\par
The rest of the paper is arranged in the following manner. In the next section we present details of the model we used and an elaborate description of the performed MC simulations. Following that in Sec.\ \ref{results} we present the results and wherever required also describe the calculations  of relevant observables. Finally, we conclude in Sec.\ \ref{conclusion} by providing a brief summary and a future outlook.

\section{Model and Method}\label{methods}
To model a binary \emph{isomeric} mixture ($A_1+A_2$) undergoing \emph{interconversion} reaction we use the nearest neighbor Ising model having the Hamiltonian
\begin{eqnarray}\label{Ising}
 \mathscr{H} = -J\sum_{\langle ij \rangle}S_iS_j,
\end{eqnarray}
where the spin $S_i=+1(\rm{or}~-1)$ represents the component $A_1 ({\rm or}~A_2)$ of the mixture, and $J~(=1)$ is the strength of interaction. The sign $\langle \dots \rangle$ denotes that only nearest-neighbor interactions are considered. We consider our system to be a square lattice ($d=2$) having a linear dimension $L$, and apply periodic boundary condition (pbc) in all directions. The Ising model in a square lattice with pbc undergoes a phase transition with a critical temperature \cite{onsager1944} 
\begin{equation}
 T_c=\frac{2J}{k_B\ln(1+\sqrt{2})},
\end{equation} 
where $k_B$ is the Boltzmann constant. For a binary mixture the critical temperature is equivalent to the demixing transition temperature.
\par
We consider that the components are undergoing the following \emph{interconversion} reaction 
\begin{equation}\label{reaction}
 A_1 \rightleftharpoons A_2. 
\end{equation}
At the same time they are also spontaneously segregating from each other, energetics of which is captured by the Hamiltonian in Eq.\ \eqref{Ising}. As already mentioned, we perform MC simulations of the above model where one needs to incorporate the dynamics of both the processes. The dynamics of the \emph{interconversion} reaction in Eq.\ \eqref{reaction} is effectively mimicked by the Glauber spin-flip move, used for capturing the nonconserved dynamics of phase ordering \cite{glauber1963,barkema_book,landau_book}. There one randomly picks up a spin on the lattice, and flips it upside down. The dynamics of segregation between the components is introduced via the Kawasaki spin-exchange move, typically used for simulating phase segregation, where the dynamics is conserved \cite{kawasaki1966,barkema_book,landau_book}. In this move, one randomly picks up a pair of nearest-neighbor spins and exchange their positions. In our simulations we attempt both the moves and accept them using the Metropolis criterion with the probability \cite{barkema_book,landau_book}
\begin{equation}
 p_i={\rm min}[1,\exp(-\Delta E/k_BT)],
\end{equation}
where $\Delta E$ is the change in energy due to the attempted move and $T$ is the temperature. For convenience we choose $k_B=1$.
\par
As the initial condition for our simulation we prepare 
a homogeneous $50:50$ binary mixture of $A_1$ and $A_2$ by placing them randomly on the lattice sites. This mimics a high-temperature disordered state. We then set the temperature to $T < T_c$ in our simulations, where we also introduce a relative weightage between the nonconserved and conserved moves using a parameter called the reaction probability $p_r$. It denotes the probability of attempting the Glauber spin-flip move at each MC step. Thus, the Kawasaki spin-exchange attempt is performed
with a probability $1-p_r$. We perform simulations for a range of values of $p_r \in [10^{-4},10^{-1}]$. We choose the unit of time to be one MC sweep (MCS), which corresponds to $L^2$ attempted MC moves. The unit of temperature is chosen to be $J/k_B$, and all the results are for a system with $L=128$ containing $L^2=16384$ number of spins or \emph{isomers}. All the simulations are run until $10^7$ MCS, which is sufficient to extract meaningful results for the chosen system size considering the covered variations of $T$ and $p_r$. Except for the time evolution snapshots all subsequently presented results are averaged over $20$ independent initial configurations, obtained by using different seeds in the random number generator.
\begin{figure*}[t!]
\centering
\includegraphics*[width=0.45\textwidth]{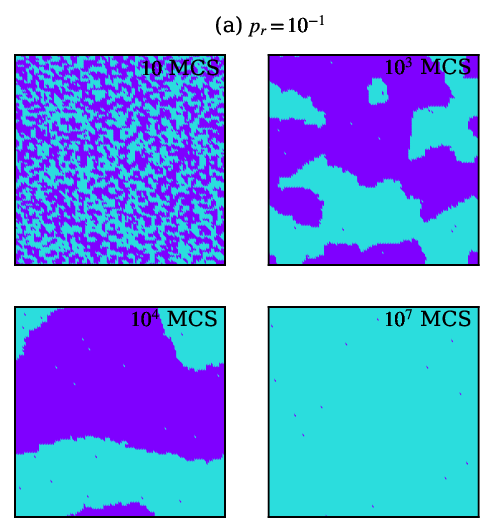}~~~~~~~~~
\includegraphics*[width=0.45\textwidth]{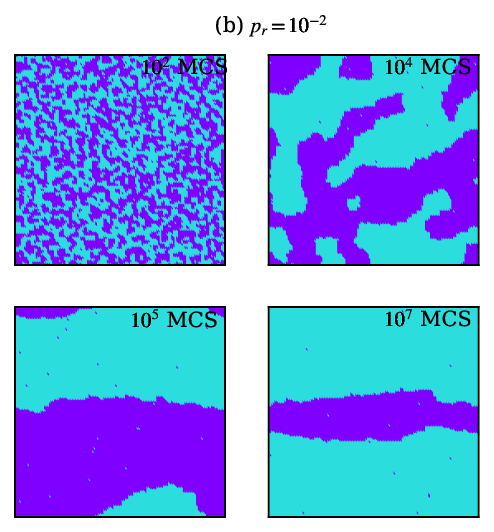}\\
\includegraphics*[width=0.45\textwidth]{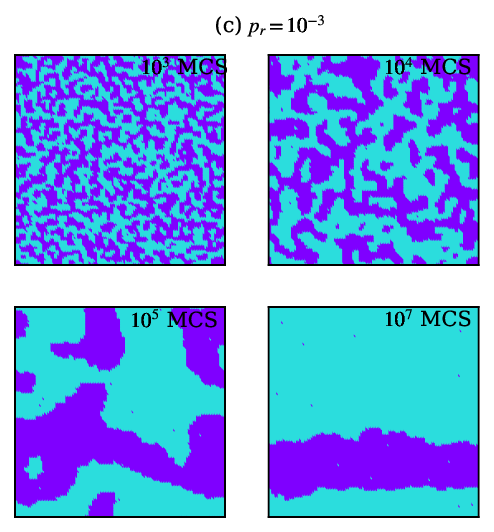}~~~~~~~~~
\includegraphics*[width=0.45\textwidth]{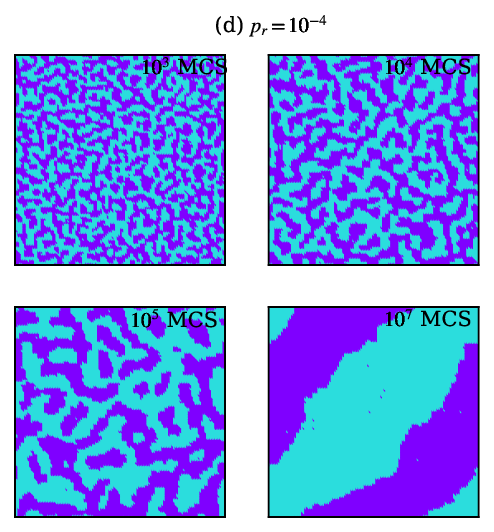}
\caption{\label{snapshots} Snapshots at different times representing the evolution of a binary mixture of \emph{isomers} ($A_1+A_2$), where an  \emph{interconversion} reaction is occurring simultaneously with segregation. The initial system is a homogeneous $50:50$ mixture of the two \emph{isomers} in a square lattice of linear dimension $L=128$, mimicking a configuration above the critical temperature $T_c$, which is then quenched to a temperature $T=0.5T_c$. Results are presented for different values of the reaction probability $p_r$, as indicated. The two colors correspond to the location of the two different \emph{isomers} on the lattice.}
\end{figure*}
\begin{figure*}[t!]
\centering
\includegraphics*[width=0.48\textwidth]{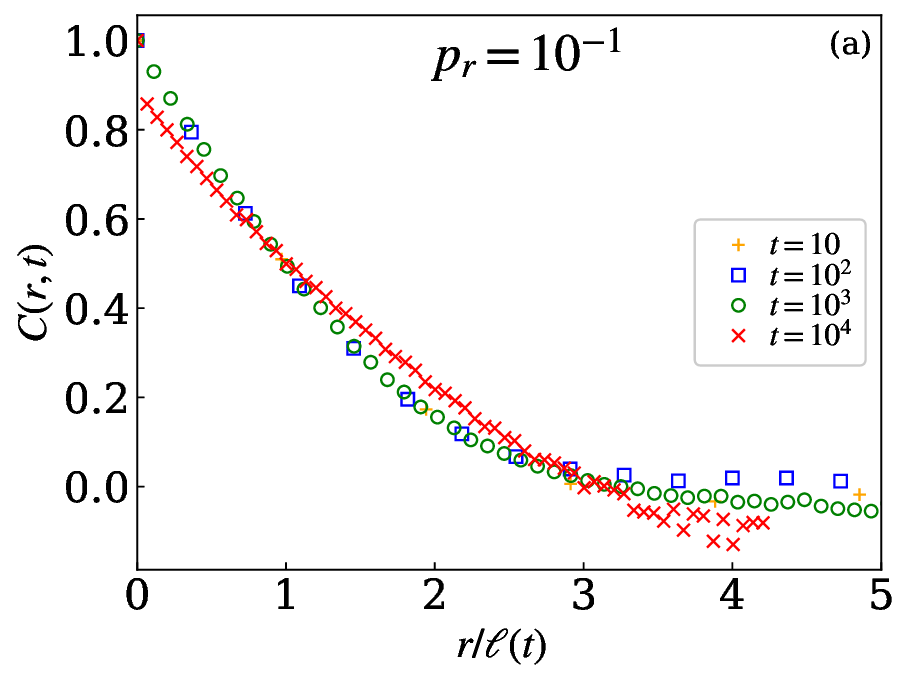}
\includegraphics*[width=0.48\textwidth]{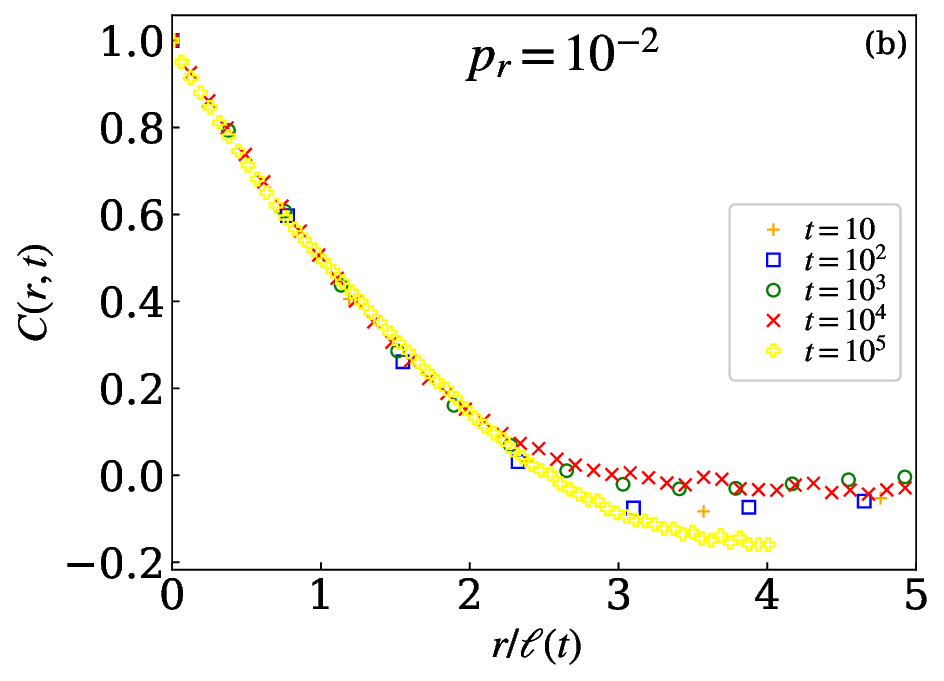}\\
\includegraphics*[width=0.48\textwidth]{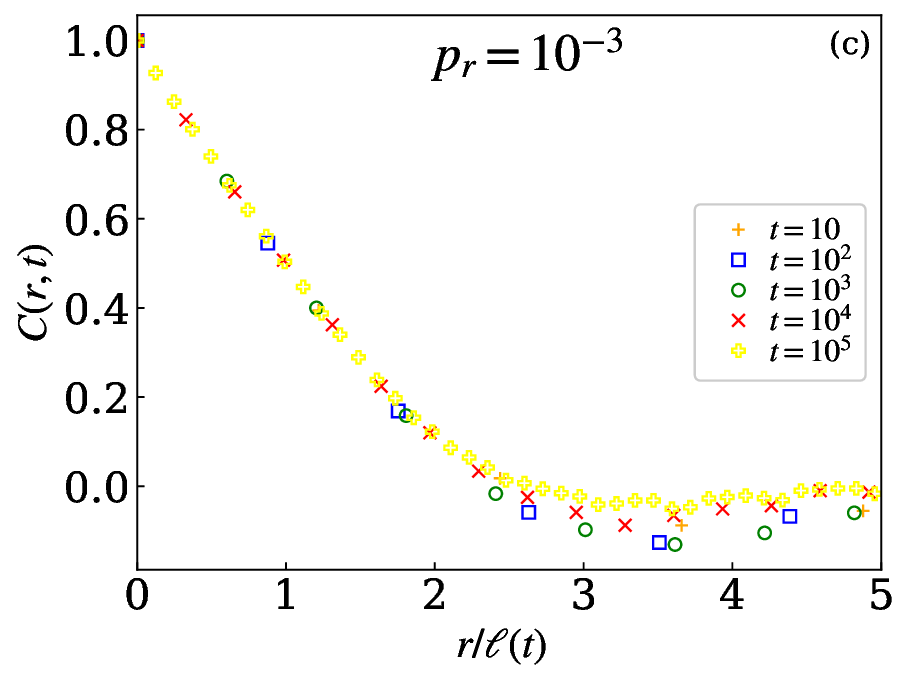}
\includegraphics*[width=0.48\textwidth]{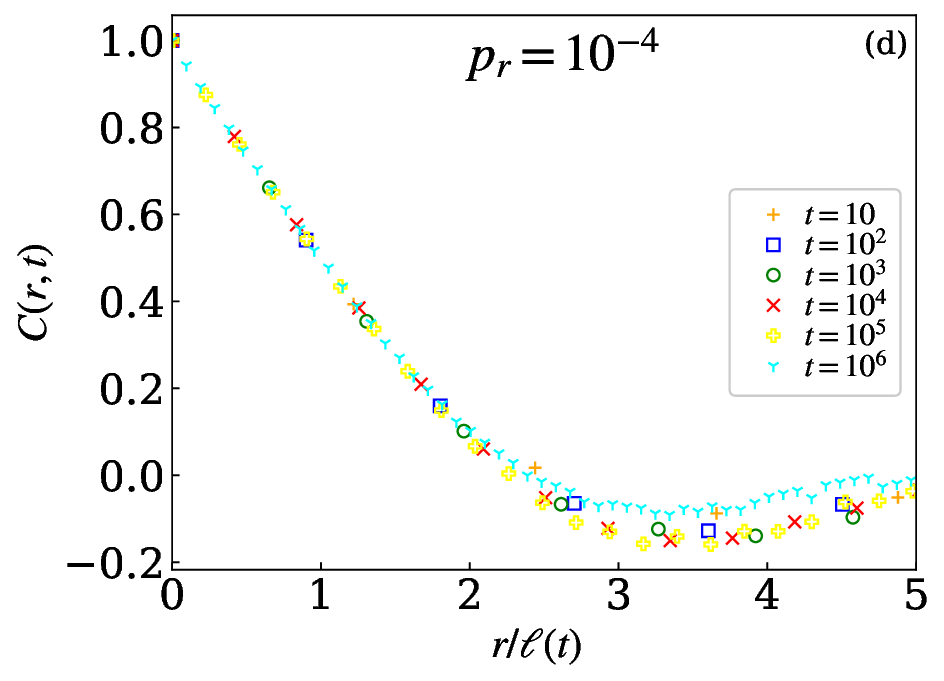}
\caption{\label{cor_func} Plots of the two-point equal-time correlation function $C(r,t)$ as a function of the scaling variable $r/\ell(t)$ at different times for different values of the reaction probability $p_r$, as indicated. Here, the average domain length $\ell(t)$ is calculate from the crossing of the unscaled data of $C(r,t)$, using the condition prescribed in Eq.\ \eqref{cr_calc}. See text for details. All results are at a temperature $T=0.5T_c$ for a system of size $L=128$. }
\end{figure*}
\section{Results}\label{results}
In this section we present our main results. It is further divided into two subsections. In the first subsection, we deal with simulation results at a fixed temperature $T=0.5T_c$. In the second, we present results obtained for variation of $T$.
\subsection{Dynamics at ${T=0.5T_c}$}\label{fixedT}
Here we present results for a fixed quench temperature $T=0.5T_c$. The effect of slow dynamics at low $T$ and difficulty of estimating various observables at high $T$ due to thermal noise are negligible at such a moderate $T$. In Figs.\ \ref{snapshots}(a)-(d) we present time evolution snapshots for different reaction probability $p_r$. As expected, the time evolution at the largest $p_r$, shown in   Fig.\ \ref{snapshots}(a), bears almost a perfect resemblance with time evolution during ferromagnetic ordering. There one can hardly notice any effect of phase segregation as the dynamics of the chemical reaction dominates throughout the evolution, and one eventually ends up with one of the \emph{isomers} as the majority or \emph{winner}, equivalent to the final magnetized state obtained during ferromagnetic ordering. As $p_r$ decreases, gradually the effects of phase segregation show up. For $p_r=10^{-3}$ and $10^{-4}$ one can clearly notice the bicontinuous domain morphology at early times, a classic signature of spinodal decomposition in a phase-segregating system \cite{marko1995,majumder2010,majumder2011}. Eventually, the reaction takes over at later times and the system approaches toward a state where one of the \emph{isomers} emerges as the \emph{winner}. For very small $p_r$ this dynamics slows down as one can appreciate from the snapshot shown at the latest time for $p_r=10^{-4}$. Since, the focus of this paper is on the domain-growth laws and its crossover dynamics, here, we do not explore the kinetics of the \emph{interconversion} reaction, and rather refer to Ref.\ \cite{Thwal1} for that.
\begin{figure*}[t!]
\centering
\includegraphics*[width=0.48\textwidth]{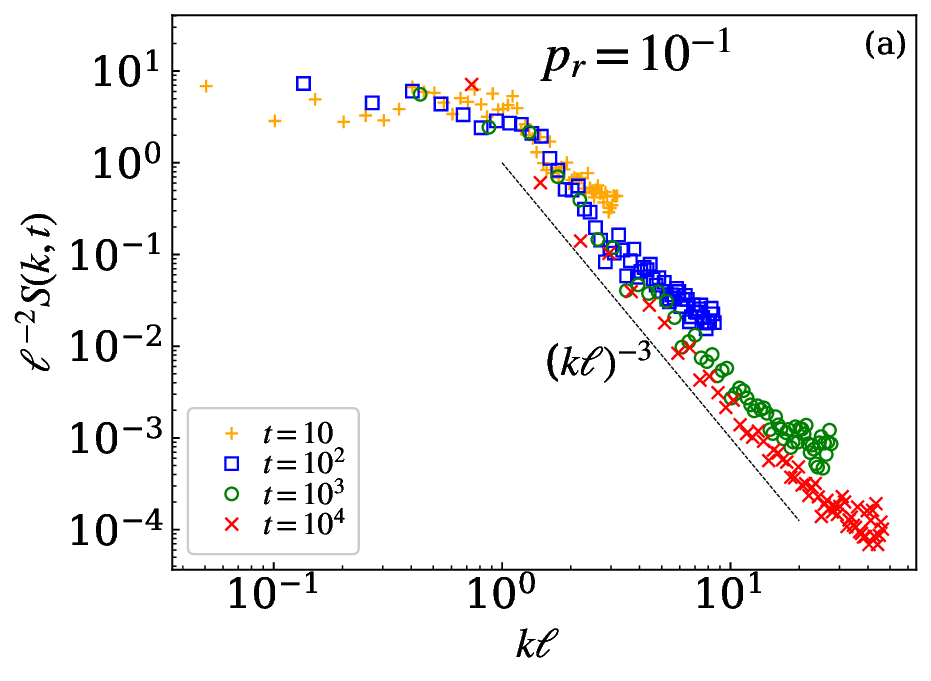}
\includegraphics*[width=0.48\textwidth]{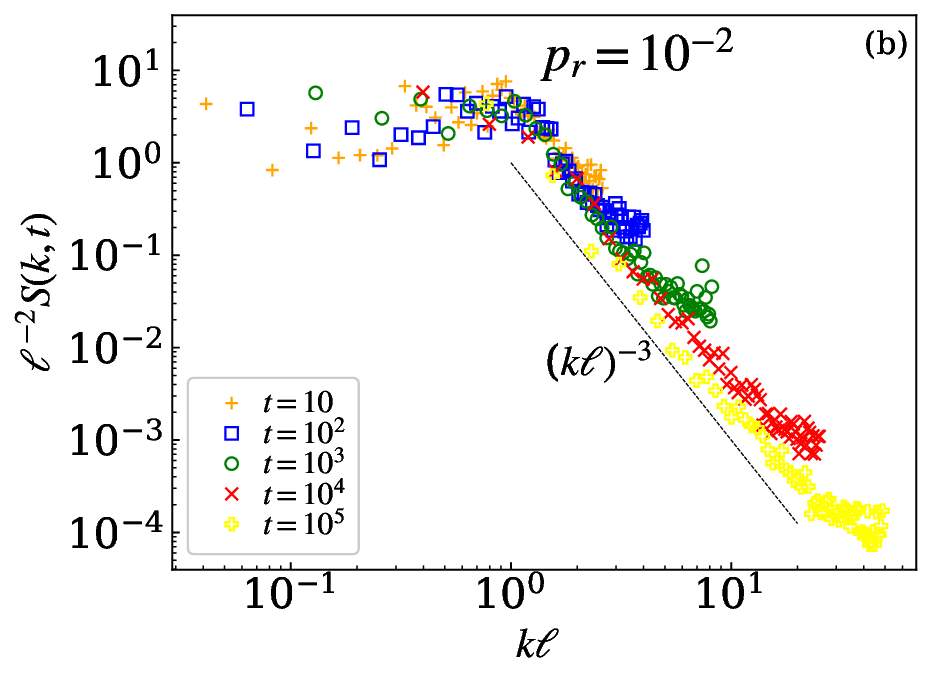}\\
\includegraphics*[width=0.48\textwidth]{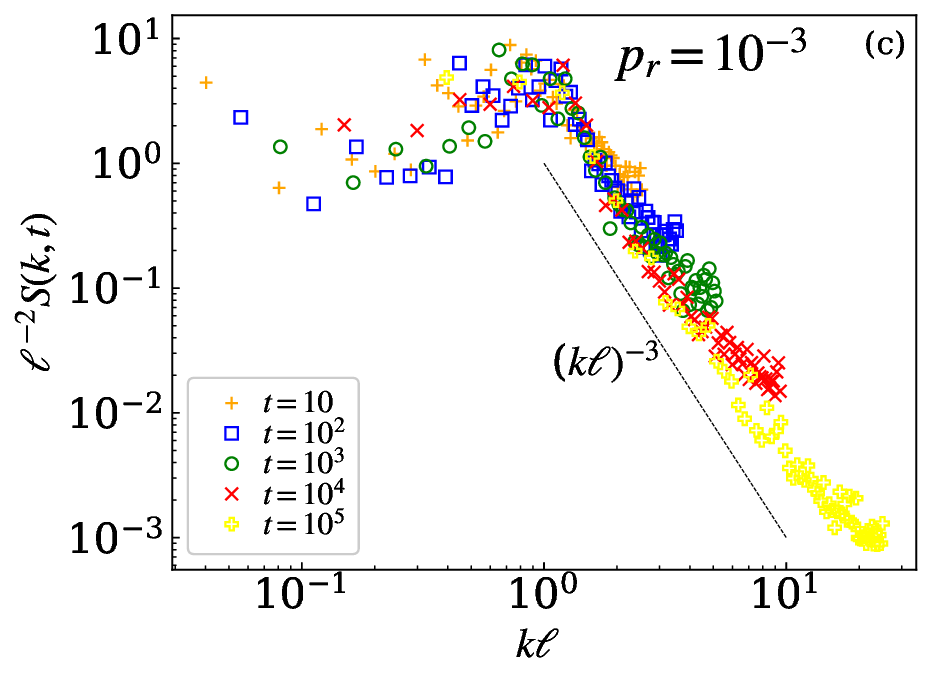}
\includegraphics*[width=0.48\textwidth]{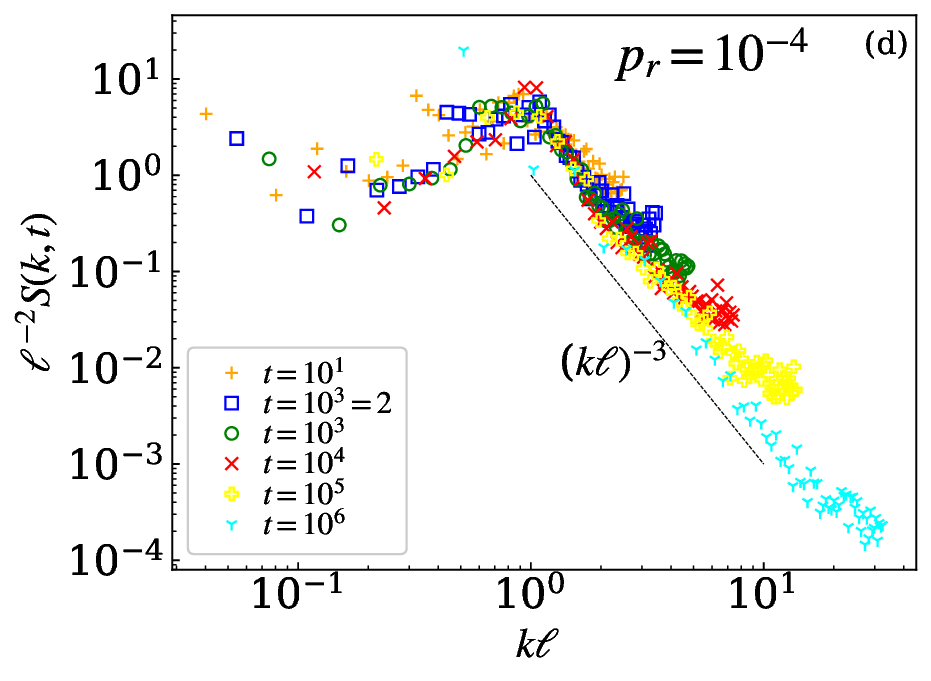}
\caption{\label{Skt} Scaling plots of the structure factor $S(k,t)$ at different times, for different values of the reaction probability $p_r$, as indicated. Here also, the average $\ell(t)$ used is obtained from $C(r,t)$. The dashed line at large $k$ represents the Porod's tail. The results are for the same systems as presented in Figs.\ \ref{cor_func}.}
\end{figure*} 
\par
It is clear from the time evolution snapshots that the pattern morphologies while the system  is evolving is affected by the two competing dynamics. As a first check we calculate the two-point equal-time  correlation function 
\begin{equation}
 C(r,t)=\langle S_iSj \rangle - \langle S_i\rangle \langle S_j\rangle,
\end{equation}
where $r=|i-j|$ is the distance between the spins at lattice sites $i$ and $j$. Following the traditional approach we extract the average domain length $\ell(t)$ using the criterion
\begin{equation}\label{cr_calc}
 C[r=\ell(t),t)]=\frac{1}{2}C(0,t).
\end{equation}
Using this $\ell(t)$, in Figs.\ \ref{cor_func}(a)-(d) we plot $C(r,t)$ against the scaled variable $r/\ell(t)$ at different times to verify the scaling quoted in Eq.\ \eqref{cor_scaling}, for different values of the reaction probability $p_r$. For the largest $p_r$, presented in Fig.\ \ref{cor_func}(a) we obtain a reasonably good collapse of data for different times until the finite size effect is experienced at $t\approx10^4$. Such a behavior is expected for ferromagnetic ordering. As $p_r$ decreases one notices that in Fig.\ \ref{cor_func}(b) the data collapse is reasonable only for intermediate times. Absence of data collapse for $t=10$ and $10^2$ suggests that the system is still under the effect of phase segregation. At the latest time the data again are affected by the finite size of the system. With further decrease of $p_r$, the data in Figs.\ \ref{cor_func}(c) and (d) show reasonably good collapse until $t=10^4$, however, with an oscillating behavior around zero at large $r$, typical of conserved dynamics of phase segregation. Note that since the dynamics for very small $p_r$ is slow, the system is not expected to experience any finite-size effects until $t\approx10^5$. Thus, the non-collapsing behavior of the data at $t=10^5$ in Figs.\ \ref{cor_func}(c) and (d) can only be attributed to the dominance of the \emph{interconversion} reaction.

\par
The scaling behavior in the kinetics can be also probed by another quantity, i.e., the structure factor which is determined by the Fourier transform 
\begin{equation}
 S(\mathbf{k},t)=\int d\mathbf{r}C(\mathbf{r},t)e^{i\mathbf{kr}}.
\end{equation}
In Figs.\ \ref{Skt}(a)-(d) we verify the scaling of $S(k,t)$, embedded in Eq.\ \eqref{skt_scaling}. Like the data for $C(r,t)$, for $p_r=10^{-1}$ the behavior of $S(k,t)$ in Fig.\ \ref{Skt}(a) is reminiscent of a typical ferromagnetic ordering. However, with decreasing $p_r$, the data presented in Figs.\ \ref{Skt}(b)-(d) show reasonably good collapse, i.e., the data cannot unambiguously capture the interplay of the conserved phase segregation dynamics and the nonconserved dynamics due to the \emph{interconversion} reaction. Noticeable is of course the power-law behavior of the tail, independent of time and $p_r$. The dashed lines in Figs.\ \ref{Skt} show the consistency of this power-law decay with the generalized Porod law for scalar order parameter given as  \cite{porod1982}
\begin{equation}
 S(k,t) \sim k^{-(d+1)}.
\end{equation}
Here the power-law exponent is $d+1=3$, mentioned next to the dashed lines. Thus, it seems that the Porod-tail behavior is unaffected by the interplay of the two dynamics.
\begin{figure}[t!]
\centering
\includegraphics*[width=0.5\textwidth]{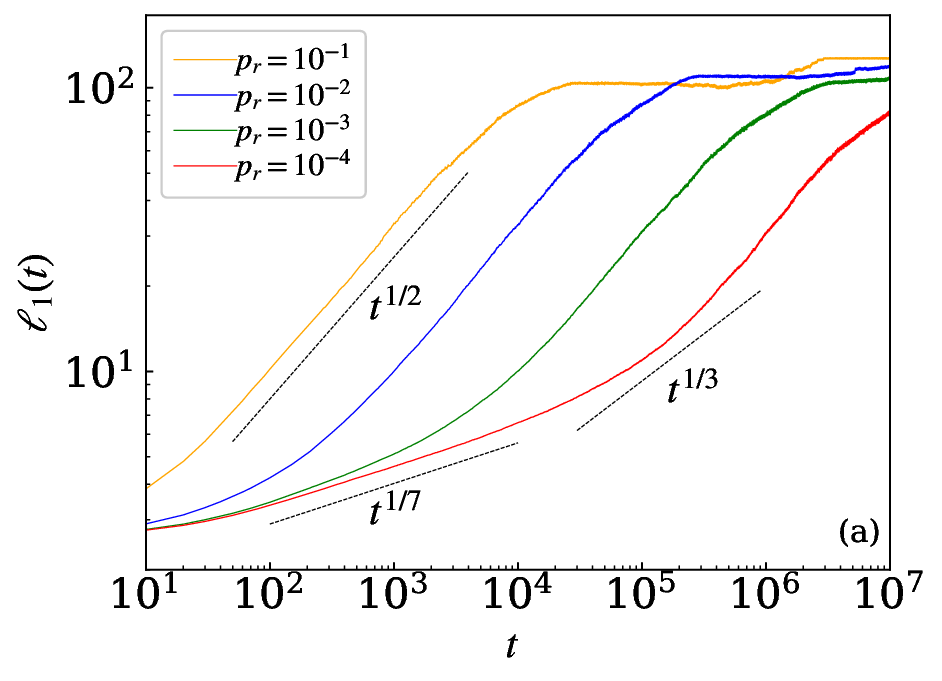}\\
\includegraphics*[width=0.5\textwidth]{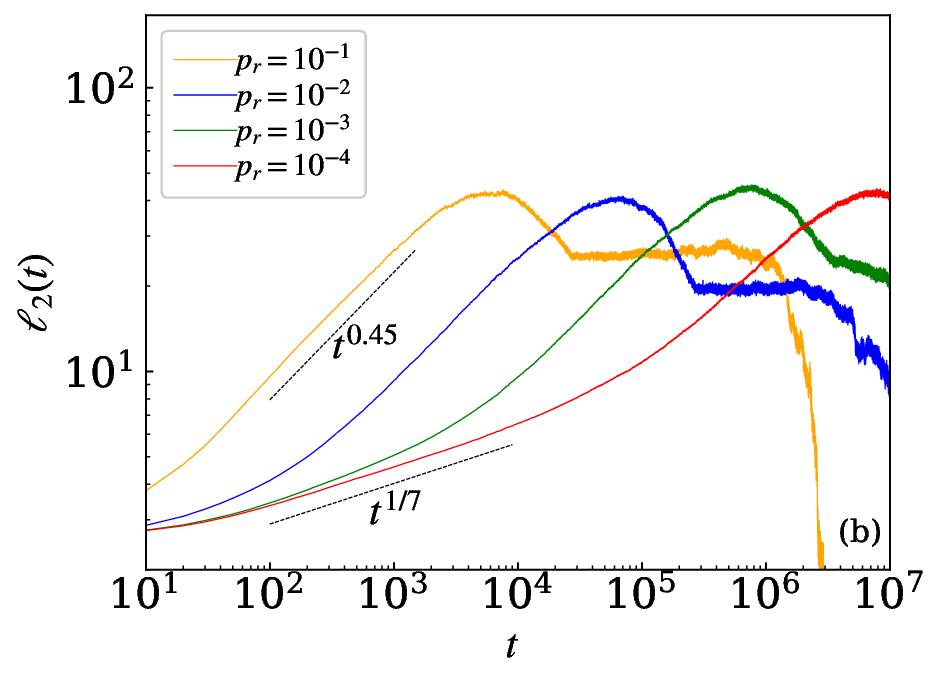}\\
\includegraphics*[width=0.5\textwidth]{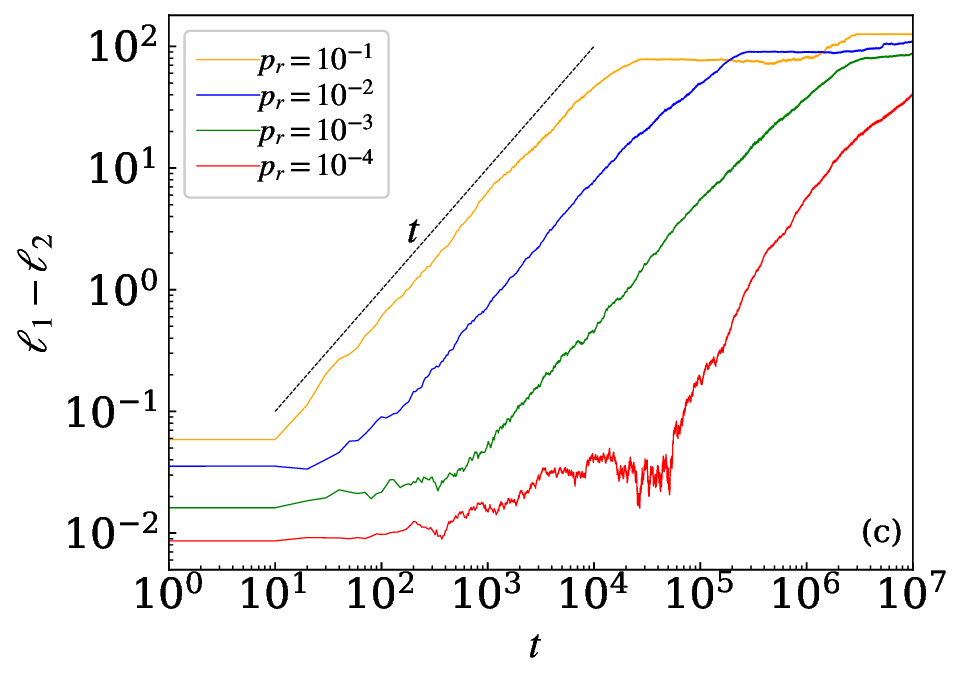}
\caption{\label{length} Double-log plots of the time dependence of the average domain size (a) of the \emph{winner} $\ell_1(t)$ and (b) the \emph{loser} $\ell_2(t)$, for different reaction probability $p_r$, obtained from simulations using a system of size $L=128$ at $T=0.5T_c$. In (c) we show the corresponding time dependence of the difference $\ell_1-\ell_2$, for different $p_r$. The dashed lines in all the figures represent different power-law behaviors as mentioned.}
\end{figure} 
\par
Next, we move on to explore the interplay of dynamics via the time dependence of the growing sizes of the domains. As shown previously by one of us that disentanglement of the kinetics of the \emph{winner} from the \emph{loser} allows one to realize the growth laws unambiguously, here also, we rely on the same \cite{majumder2023}. Thus, rather than using the average domain size $\ell(t)$, at a given time we estimate the average domain size $\ell_1(t)$ of the \emph{winner} and also $\ell_2(t)$ of the \emph{loser}. For a detail on how these lengths are estimated, we refer to Ref.\ \cite{majumder2023}. The measured domain lengths of the \emph{winner} for different $p_r$ are presented in Fig.\ \ref{length}(a) on a double-log scale. The data for $p_r=10^{-1}$ show a single growth regime, consistent with a $\ell_1(t) \sim t^{1/2}$ power-law behavior, i.e., the LCA law. The flat behavior following the power-law growth indicates that finite-size effects have started showing up. At even later times, one sees a jump that correspond to an \emph{avalanche} \cite{olejarz2013,majumder2023}. As $p_r$ decreases an initial regime with a much slower growth emerges, by virtue of the dominance of phase segregation dynamics over the \emph{interconversion} reaction dynamics. At later times, one sees again a growth consistent with the LCA law. The duration of the initial slow regime gets extended as $p_r$ decreases. In fact for $p_r=10^{-4}$, within the given maximum simulation time, the slowest regime  appears to be longer lived.  The data for $p_r=10^{-3}$ follow the data for $p_r=10^{-4}$ until $t \approx 500$, when the reaction dynamics takes over. The data for $p_r=10^{-4}$ continues to grow slowly, consistent with a $\ell_1 \sim t^{1/7}$ behavior, which is much slower than the expected LS behavior $\ell_1 \sim t^{1/3}$ for a phase-segregating system. This implies that the growth of the domains slows down  significantly in presence of an \emph{interconversion} reaction. On the other hand, when the domains of the two isomers are well separated, phase segregation seizes and the dynamics is entirely controlled by the \emph{interconversion} reaction. Hence, at late times the LCA behavior is always realized as shown by the data for all $p_r$.

\par
In Fig.\ \ref{length}(b) we show the time dependence of the average domain size $\ell_2(t)$ of the \emph{loser} for different $p_r$. There also for $p_r=10^{-1}$ one sees a single growth regime followed by a plateau before it suddenly vanishes due to the \emph{avalanche} effect, also observed as a corresponding abrupt increase of $\ell_1(t)$ in Fig.\ \ref{length}(a). In the power-law regime, the growth is consistent with $\ell_2(t) \sim t^{0.45}$, albeit, slower than what is observed for the corresponding $\ell_1(t)$. With decreasing $p_r$ a slower early-time regime emerges like in the time dependence of $\ell_1(t)$. Here, also the  early-time regime gets extended as $p_r$ decreases. For $p_r=10^{-4}$ the growth in this early-time regime is consistent with a $\ell_2(t) \sim t^{1/7}$ behavior, similar to what is observed for the corresponding $\ell_1(t)$ in Fig.\ \ref{length}(a). This indeed confirms that in this early-time regime the dynamics is controlled by the phase segregation, however, the occurrence of the \emph{interconversion} reaction every now and then makes  the dynamics slower than the LS growth, expected for an ideally phase-segregating system. For smaller $p_r$ the very late-time behavior is not observed within the given maximum simulation time. Overall, from the plots presented in Figs.\ \ref{length}(a) and (b) we infer that during the evolution initially the phase segregation is dominant with occasional involvement of the \emph{interconversion} reaction, and thus the growth of the domains of the \emph{isomers} in this regime is even slower than LS law. However, at later times phase segregation is no more capable of driving the system, as either the system has completely phase segregated or the domains of the individual \emph{isomers} are very well separated. Hence at reaction-dominated late times, the dynamics is perfectly consistent with the LCA law, expected for ferromagnetic ordering.
\begin{figure}[t!]
\centering
\includegraphics*[width=0.48\textwidth]{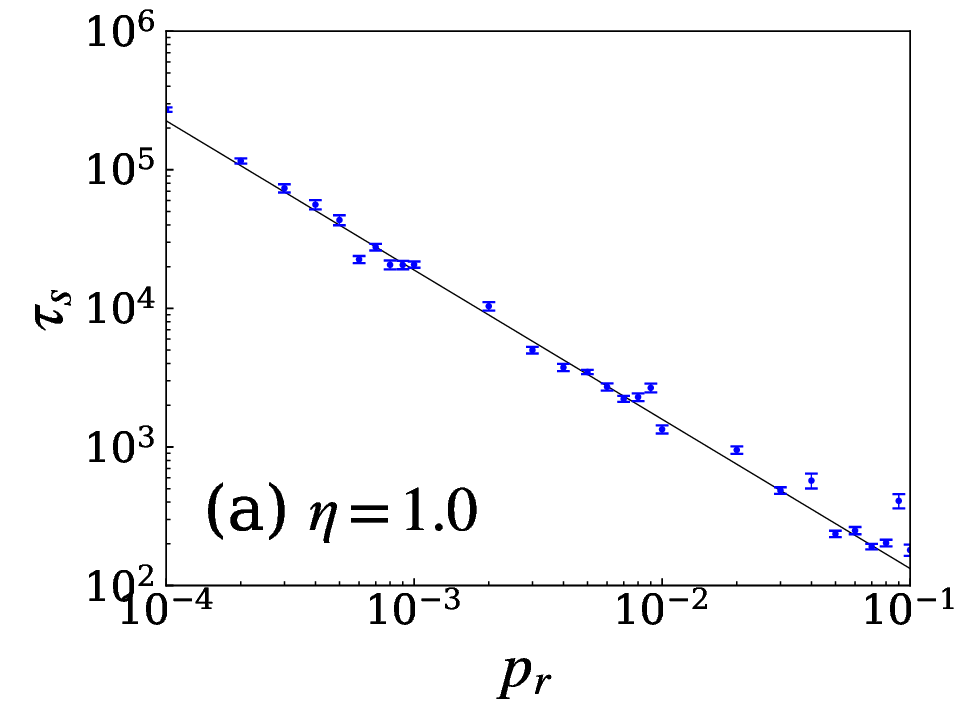}\\
\includegraphics*[width=0.48\textwidth]{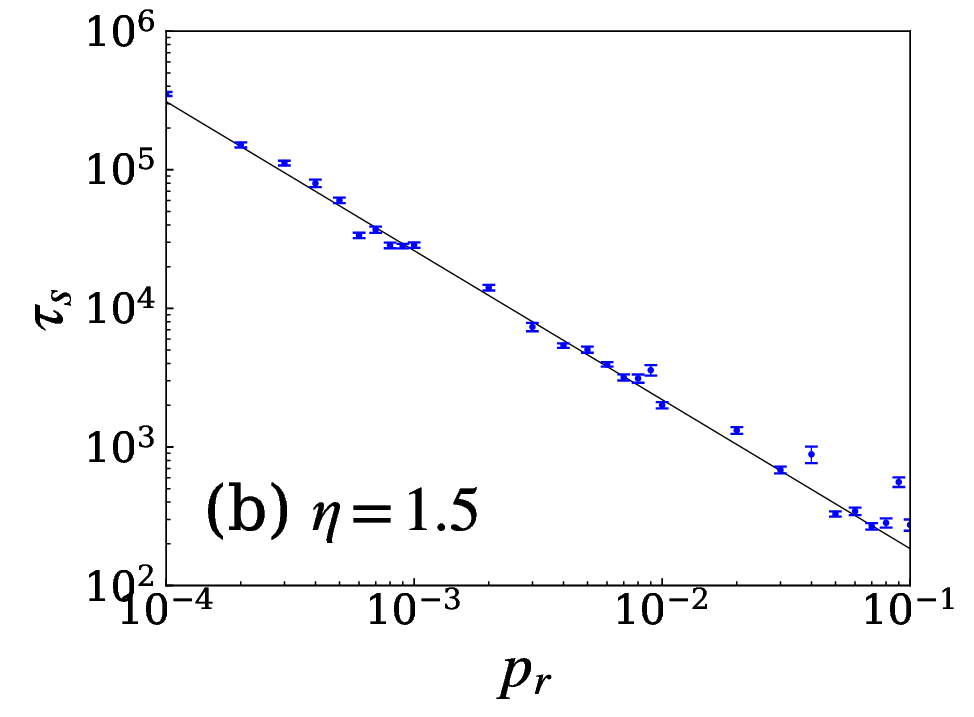}\\
\includegraphics*[width=0.48\textwidth]{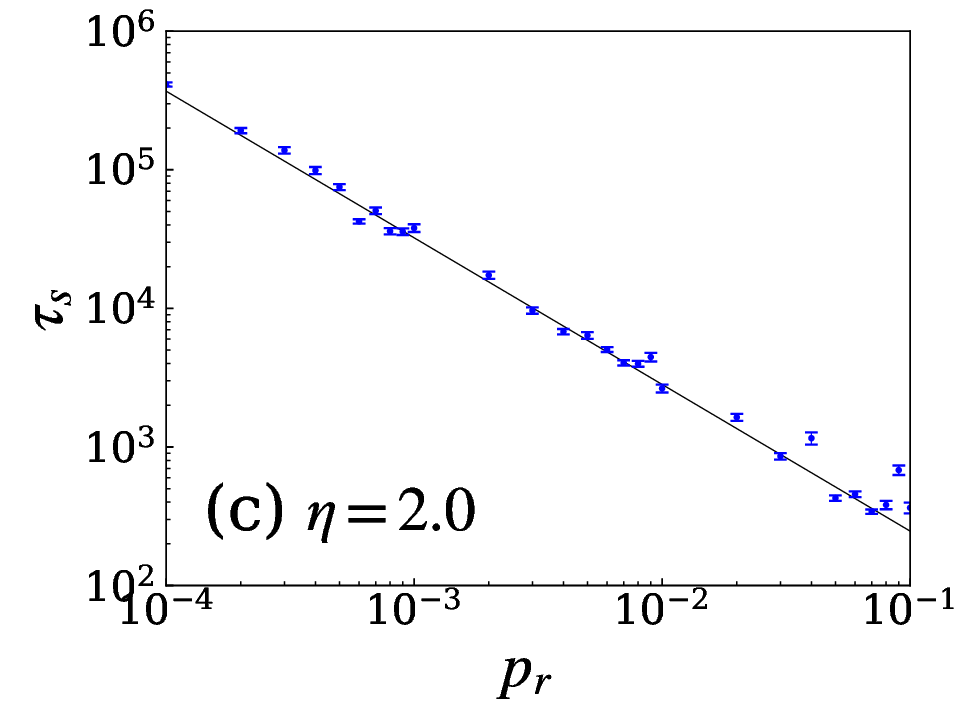}
\caption{\label{taus_vs_pr} Double-log plots of the crossover time $\tau_s$ as a function of the reaction probability $p_r$, at $T=0.5T_c$ using a system of size $L=128$, 
for different choices of the parameter $\eta$ in Eq.\ \eqref{eta_eqn}. The dashed lines are the best fits obtained using the ansatz in Eq.\ \eqref{taus_fit}. For details see the main text.}
\end{figure}

\par
The other important aspect of the kinetics is of course the time scale when the system crosses over to the reaction dominated regime. This crossover is more clearly visible if one plots the time dependence of the difference between the domain length of the \emph{winner} and \emph{loser}, i.e., $\ell_1-\ell_2$, as presented in Fig.\ \ref{length}(c) for the data in Figs.\ \ref{length}(a) and (b). In this case the behavior is almost similar for all $p_r$. Initially for a certain period, $\ell_1-\ell_2$ remains constant followed by a steady growth consistent with the dashed line representing
\begin{equation}
 \ell_1-\ell_2 \sim t,
\end{equation}
i.e., linear behavior with time. It can be easily noticed that the time when the flat behavior of $\ell_1-\ell_2$ switches over to a linear growth, shifts toward right with decrease in $p_r$. From this time onward the growth of $\ell_1$ is significantly faster than the corresponding growth in $\ell_2$, thus indicating a crossover to the reaction dominated dynamics. Based on this observation we define the crossover time $\tau_s$ from segregation to reaction dominated regime such that it satisfies the relation 
\begin{equation}\label{eta_eqn}
 \ell_1(\tau_s)-\ell_2(\tau_s)=\eta,
\end{equation}
where $\eta$ prescribes how much difference between $\ell_1$ and $\ell_2$ is considered. In Figs.\ \ref{taus_vs_pr}(a)-(c) we show the plots of $\tau_s$ as a function of $p_r$ for three different choices of the parameter $\eta$ in Eq.\ \eqref{eta_eqn}. There the errors are estimated from a
Jackknife analysis, where $\tau_s$ is calculated  independently for each Jackknife bin that contains all but data from one of the $20$ initial realizations we used \cite{efron1982}. The apparent linear behavior of the data on a double-log scale implies a power-law dependence of $\tau_s$ on $p_r$. To quantify this power-law behavior we write down the following ansatz
\begin{equation}\label{taus_fit}
 \tau_s=Ap_r^{-x},
\end{equation}
where $A$ is a prefactor and $x$ is the power-law exponent. We fit this ansatz to our data, the results of which are tabulated in Table\ \ref{tab1}. In Figs.\ \ref{taus_vs_pr}(a)-(c), the dashed lines represent the obtained best fits. The results suggest that although the prefactor $A$ increases with $\eta$, the estimated exponent $x\approx1.07$ is very weakly dependent on the choice of $\eta$. Hence, for all subsequent estimation of $\tau_s$ we rely only on the choice $\eta=1$.  
 \begin{table}[t!]
 \caption{Fitting results for $\tau_s$ vs. $p_r$ data presented in Figs.\ \ref{taus_vs_pr}, using the ansatz in Eq.\ \eqref{taus_fit}.}
\begin{tabular}{m{1.0cm} m{1.75cm} m{1.5cm}} 
 \hline
 \hline
 ~~$\eta$ & ~~~~~~~$A$ & ~~~~~~$x$ \\
 \hline
 \hline
  ~$1.0$& $11.12\pm1.1$ & $1.08\pm 0.02$ \\  
  ~$1.5$& $15.53\pm1.6$ & $1.08\pm 0.02$ \\ 
  ~$2.0$& $21.63\pm2.1$ & $1.06\pm 0.02$ \\ 
  \hline
\end{tabular}
\label{tab1}
\end{table}
\begin{figure*}[t!]
\centering
\includegraphics*[width=0.48\textwidth]{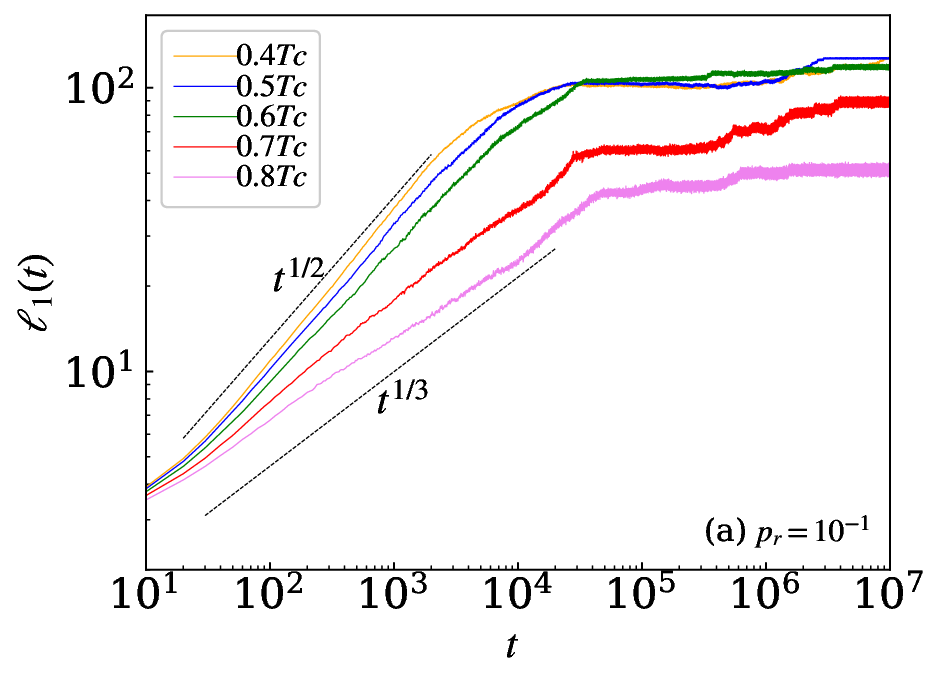}
\includegraphics*[width=0.48\textwidth]{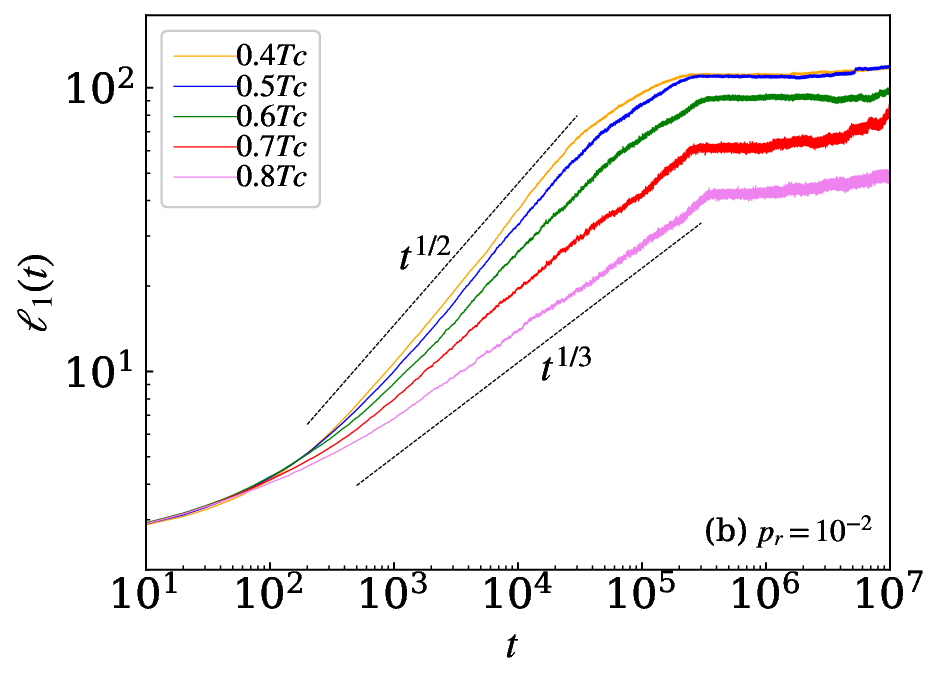}\\
\includegraphics*[width=0.48\textwidth]{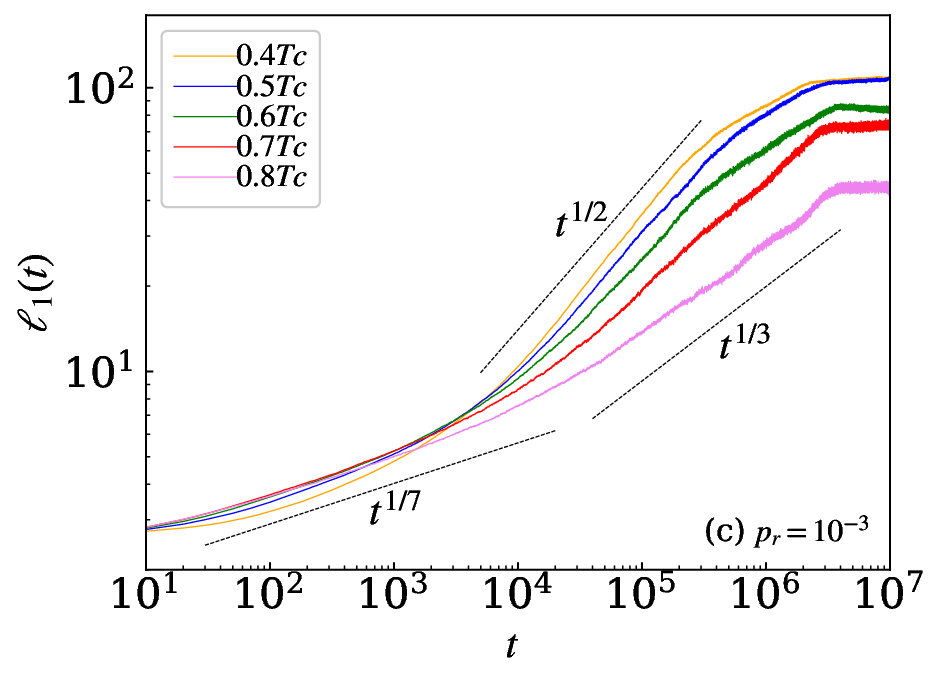}
\includegraphics*[width=0.48\textwidth]{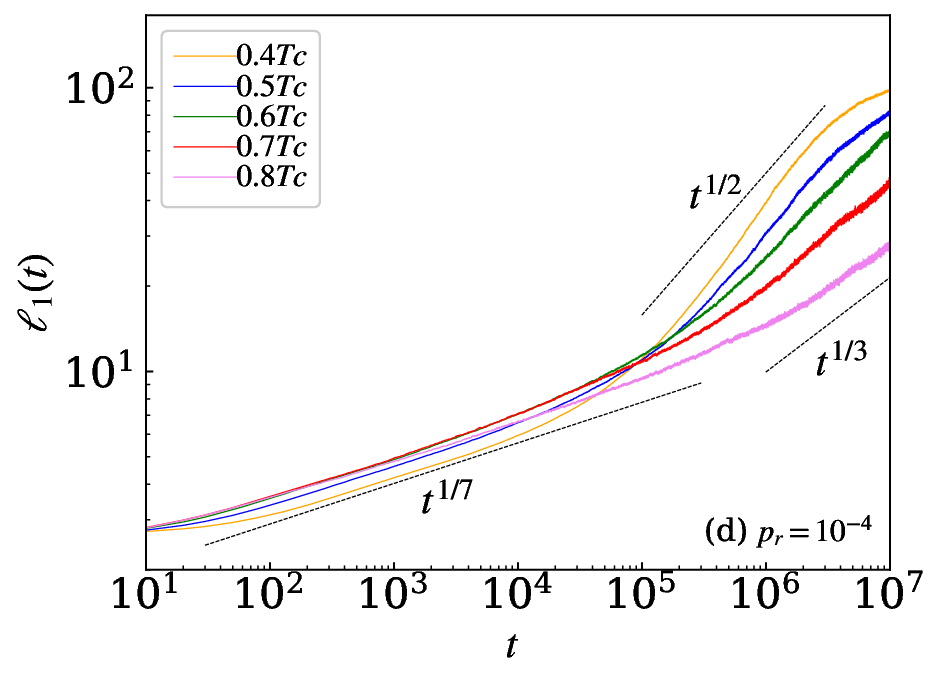}
\caption{\label{winner_diffT} Double-log plots of the time dependence of the average domain size of the \emph{winner} $\ell_1(t)$ at different temperatures for different reaction probability $p_r$, using a system of size $L=128$. The two dashed lines representing power laws $\ell_1 \sim t^{1/2}$ and $\ell_1\sim t^{1/3}$ are respectively, the LCA and LS growths. In (c) and (d) an additional dashed line representing the power-law $\ell_1 \sim t^{1/7}$ is also shown for the initial regime.}
\end{figure*}
 \subsection{Temperature dependence of the dynamics}
 In this subsection we investigate the temperature dependence of the kinetics. For that purpose we perform simulations for different values of the reaction probability $p_r$, at different temperatures below $T_c$.
 \par
 Here, we do not present the time evolution snapshots for different temperatures, and rather straight away move on to the time dependence of the average domain sizes. We start with plots of the average domain size $\ell_1(t)$ of the \emph{winner} in Figs.\ \ref{winner_diffT}, for five different temperatures. For the largest reaction probability, i.e., $p_r=10^{-1}$, presented in Fig.\ \ref{winner_diffT}(a), the growth of $\ell_1(t)$ consists of a single regime for all $T$. At moderate temperatures, i.e., $T=0.4T_c$, $0.5T_c$ and $0.6T_c$, this growth seem to be consistent with the LCA law $\ell_1 \sim t^{1/2}$. However, for higher $T$, significant deviation from the LCA law is observed as the growth becomes slower with increase in $T$. The inconsistency of the data with the other dashed line implies that the slow growth is certainly faster than the LS growth $\ell_1(t)\sim t^{1/3}$. Also, at high $T$ there is no chance of dynamic freezing \cite{spirin2001}. Thus, the apparent slow growth can be attributed to the presence of significantly large noise clusters at high $T$ that does not allow estimation of $\ell_1(t)$ from a pure domain morphology of the \emph{winner}. This reasoning can further be appreciated from the fact that at very late time when the reaction has almost finished, $\ell_1(t)$ is supposed to be saturated at a value $\approx L$, at high $T$ this happens at $\ell_1(t) < L$. This problem can of course be tackled by appropriate noise removal technique \cite{majumder2010,majumder2011,majumder2013}. Nevertheless, we abstain ourselves from doing so and rather stress that the observation of a single growth regime independent of $T$ suggests a reaction dominated dynamics for $p_r=10^{-1}$ at all $T$. 
 \begin{figure*}[t!]
\centering
\includegraphics*[width=0.48\textwidth]{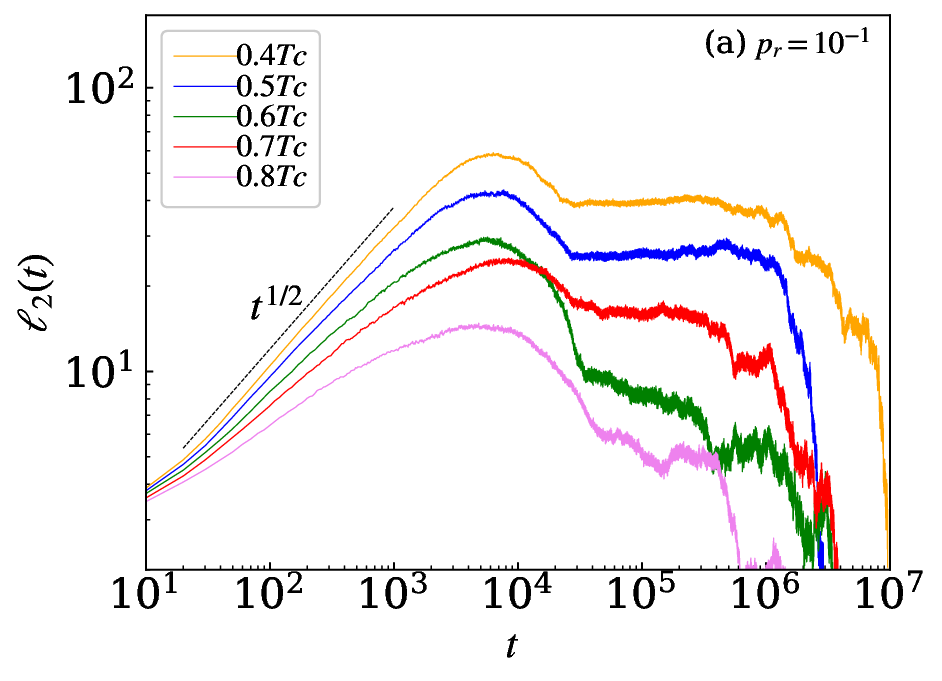}
\includegraphics*[width=0.48\textwidth]{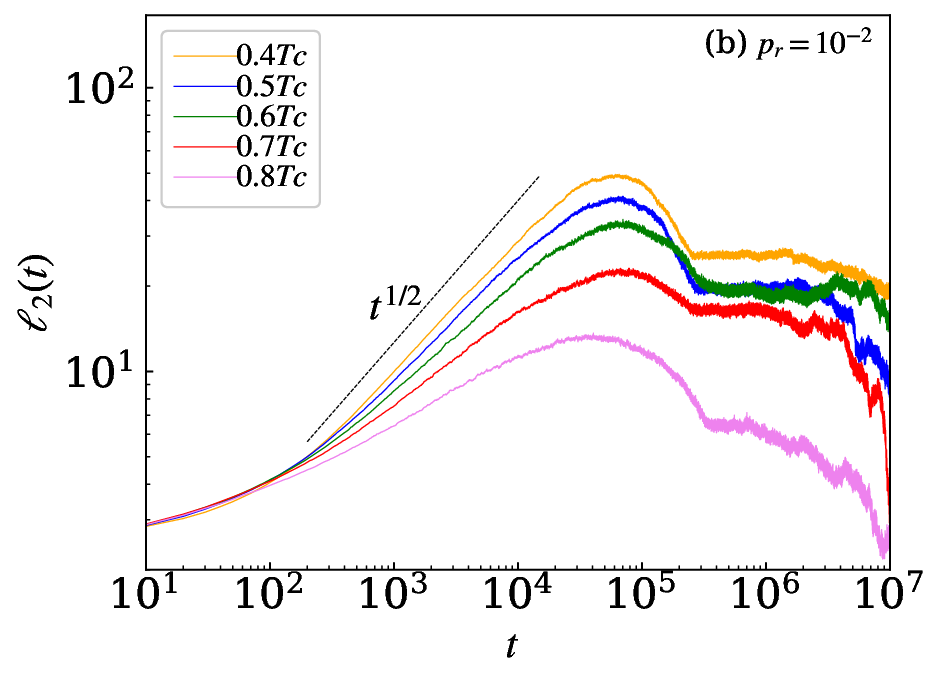}\\
\includegraphics*[width=0.48\textwidth]{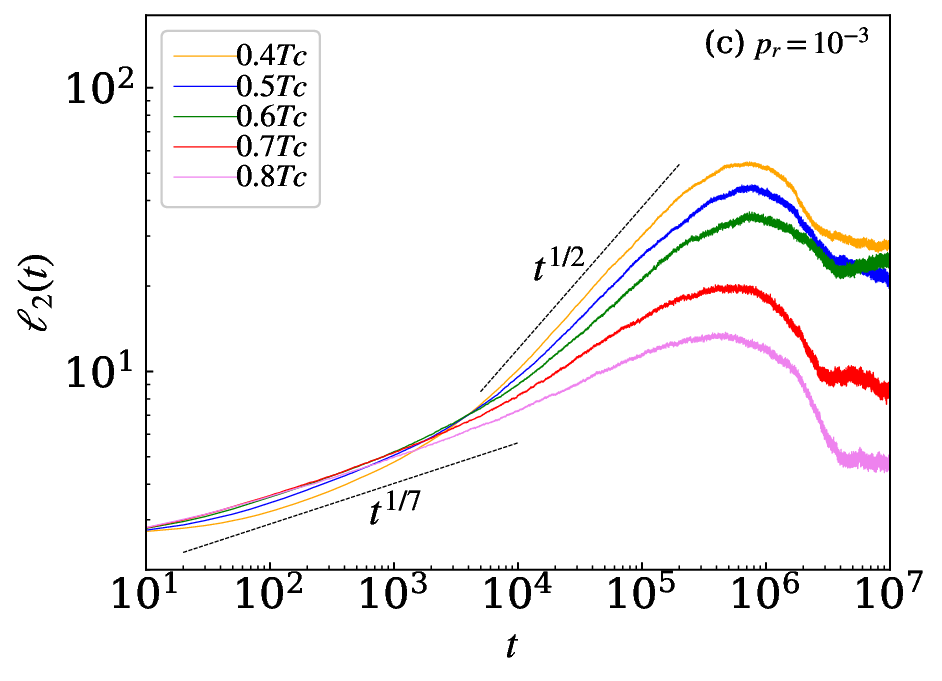}
\includegraphics*[width=0.48\textwidth]{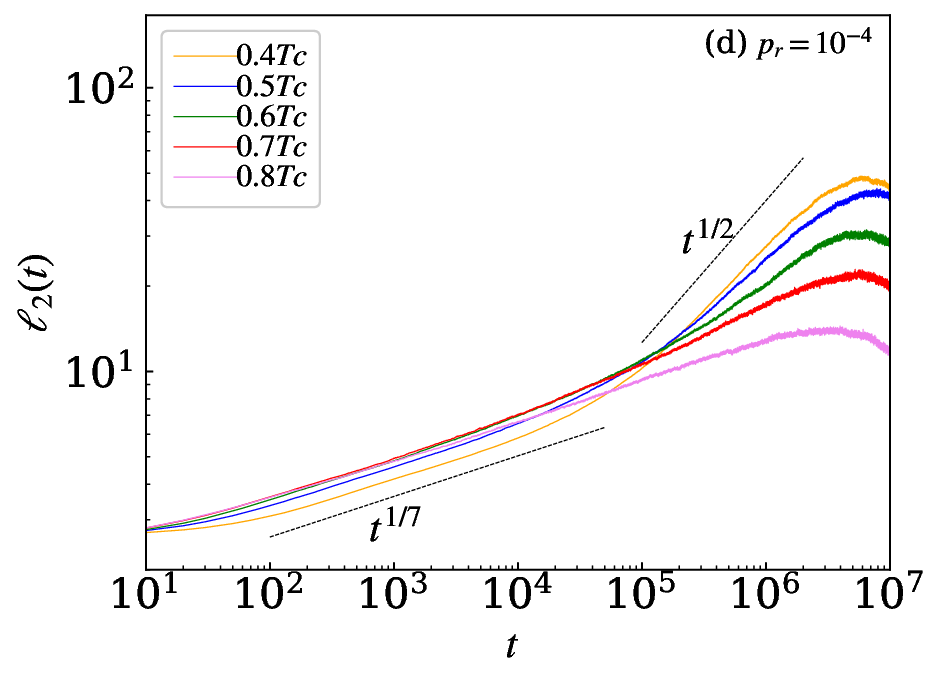}
\caption{\label{loser_diffT} Double-log plots of the time dependence of the average domain size of the \emph{loser} $\ell_2(t)$ at different temperatures for different reaction probabilty $p_r$, using a system of size $L=128$. The dashed line representing power law $\ell_2 \sim t^{1/2}$ is for the LCA growth. In (c) and (d) $\ell_2(t)\sim t^{1/7}$ for the initial regime is also shown as a dashed line.}
\end{figure*}
 \par
 As $p_r$ decreases one can clearly see the emergence of an initial slow-growth regime for $\ell_1(t)$ at all temperatures. As shown in Fig.\ \ref{winner_diffT}(b), for $p_r=10^{-2}$ the initial slow-growth regime is very short lived as there still the reactions start to control the dynamics from early time. Followed by the slow growth, $\ell_1(t)$ shows a faster growth that is again consistent with the LCA law at moderate temperatures. Again, for higher temperatures the growth although faster than the LS growth, is apparently significantly slower than the LCA law which again could be atrributed to the effect of noisy clusters while estimating $\ell_1(t)$. 
 \par
 For even smaller $p_r$, presented in Figs.\ \ref{winner_diffT}(c) and (d), the initial slow regime of $\ell_1(t)$ is long lived and prominent. Consistency of the data at all $T$ with the power-law behavior $\ell_1 \sim t^{1/7}$ suggests its robustness. At later times the data crosses over to a faster growth when the dynamics is dominated by the \emph{interconversion} reaction. The growth then again is consistent with the LCA law for moderate $T$. At high $T$, for both $p_r=10^{-3}$ and $10^{-4}$ the data again show significant deviation from the LCA law. For $p_r=10^{-4}$ at $T=0.8T_c$ the data even seem to be consistent with a $\ell_1(t)\sim t^{1/3}$ behavior. However, we stress that this is just a mere coincidence and should not be treated as the realization of the LS behavior. In fact, this slow growth at late time as already mentioned is an artifact of the estimation of $\ell_1(t)$ from impure domain morphology at high $T$.
 \par
 The temperature dependence of the average domain size of the \emph{loser} $\ell_2(t)$ corresponding to the cases discussed above are presented in Figs.\ \ref{loser_diffT}. For the largest value of $p_r=10^{-1}$, shown in Fig.\ \ref{loser_diffT}(a), the comparative behavior among different $T$ is similar to what is observed for $\ell_1(t)$ in Fig.\ \ref{winner_diffT}(a). At the lowest $T$ the data is consistent with the LCA law. At later times the data show a flat behavior before eventually they decay and almost vanish. As $T$ increases the behavior is similar, except for the fact that the growth becomes slower, which again is due to the fact that the estimated $\ell_2(t)$ is not from a pure domain morphology. Noticeable is that the time taken by $\ell_2(t)$ to almost vanish, decreases as $T$ increases. This is a signature of the fact that for $p_r=10^{-1}$, the total reaction time or kinetics of the reaction follows an Arrhenius behavior \cite{Thwal1}. 
 
 \par
 The behavior of the data for $p_r=10^{-2}$, presented in Fig.\ \ref{loser_diffT}(b) is similar to what is observed for $p_r=10^{-1}$. At low $T$, in the growth regime again the data are almost consistent with the LCA law, and deviate significantly as $T$ increases. Following a brief period of flat behavior, $\ell_2(t)$ start to decay and almost vanishes at late times. The trend of the data again indicates an Arrhenius behavior. With further decrease of $p_r$, within the growth regime there exists two sub regimes as shown by the plots presented in Figs.\ \ref{loser_diffT}(c) and (d), like what is observed for $\ell_1(t)$ in Figs.\ \ref{winner_diffT}(c) and (d). The initial slow-growth regime here is consistent with the power law $\ell_2(t) \sim t^{1/7}$, similar to how $\ell_1(t)$ behaves. This confirms that irrespective of $T$, when the reaction probability $p_r$ is small, initially the dynamics is almost conserved due to the dominant effect of segregation, however, it is much slower than the LS growth. At later times when the \emph{interconversion} reaction starts to dominate the growth becomes faster for a brief period, albeit slower than the LCA growth, before eventually starting to decay. Within the maximum simulation time of $10^7$ MCS, for $p_r=10^{-3}$ and $10^{-4}$ the data for $\ell_2(t)$ do not reach the point where they almost vanish. Hence, from this data it is not possible to infer anything about the Arrhenius behavior of the \emph{interconversion} reaction. In this regard we refer to Ref.\ \cite{Thwal1} where it has been shown that for lower $p_r$ values, the Arrhenius behavior of the \emph{interconversion} reaction gets disrupted.
 \begin{figure}[t!]
\centering
\includegraphics*[width=0.5\textwidth]{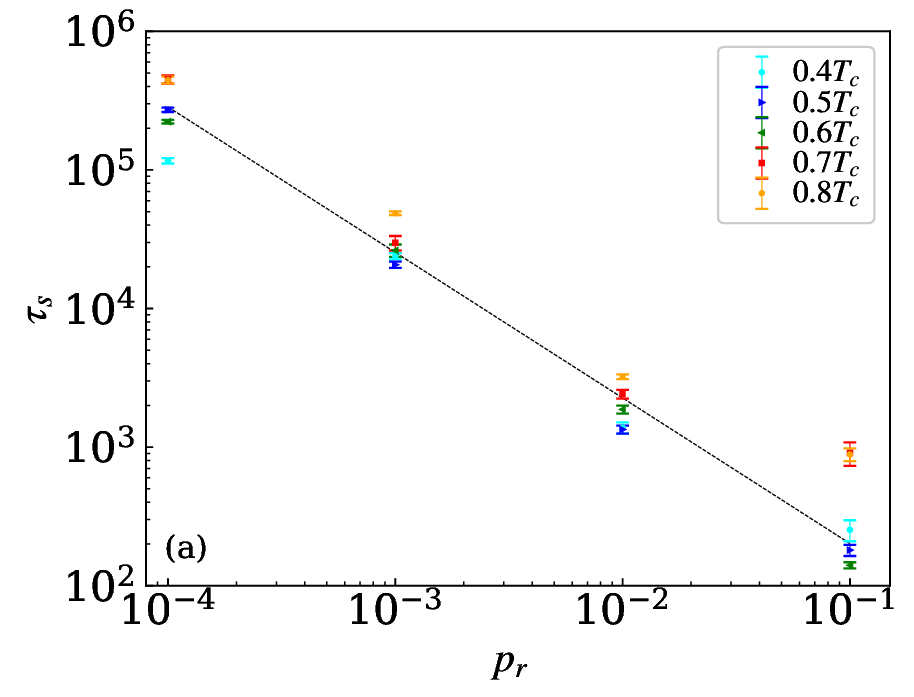}\\
\includegraphics*[width=0.5\textwidth]{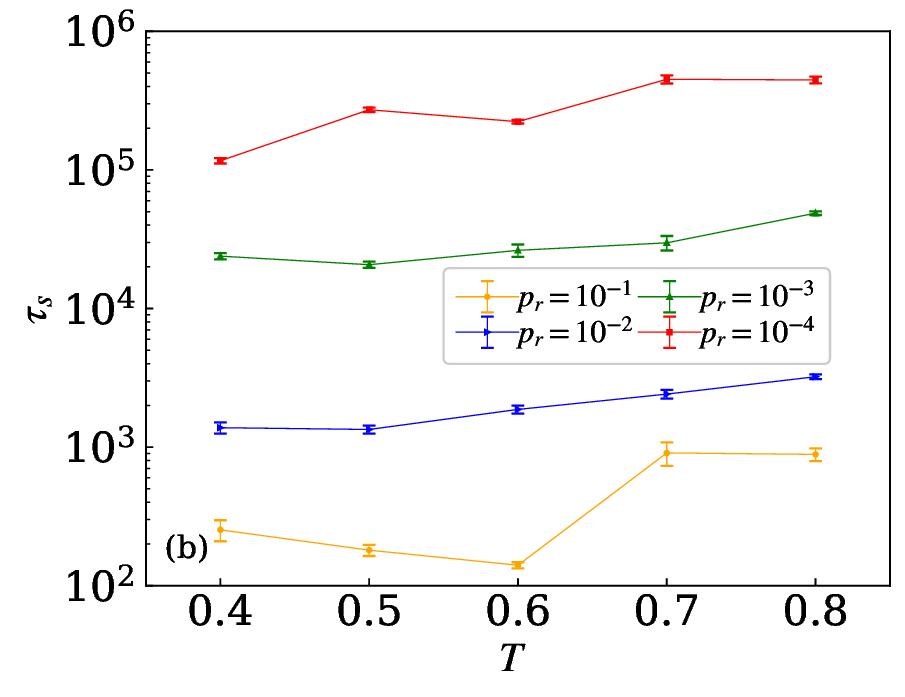}
\caption{\label{taus_diffT_pr} (a) Double-log plots of the crossover time $\tau_s$ as a function of the reaction probability $p_r$, at different $T$ using a system of size $L=128$. The dashed lines represents Eq.\ \eqref{taus_fit} with $A=17.97$ and $x=1.05$. (b) Variation of $\tau_s$ as a function of $T$, for fixed $p_r$ as mentioned.}
\end{figure}
\begin{table}[b!]
 \caption{Fitting results for $\tau_s$ vs. $p_r$ data at different temperatures using the ansatz in Eq.\ \eqref{taus_fit}.}
\begin{tabular}{m{1.25cm} m{2.5cm} m{1.5cm}} 
 \hline
 \hline
 ~~~$T$ & ~~~~~~~~~$A$ & ~~~~~~~~$x$ \\
 \hline
 \hline
  ~$0.4T_c$& $26.54\pm18.8$ & $0.92\pm 0.08$ \\  
%   ~$0.5T_c$& $11.12\pm1.6$ & $1.08\pm 0.02$ \\ 
  ~$0.6T_c$& $12.74\pm1.7$ & $1.06\pm 0.02$ \\ 
  ~$0.7T_c$& $13.64\pm6.4$ & $1.13\pm 0.06$ \\
  ~$0.8T_c$& $25.81\pm12.8$ & $1.07\pm 0.07$ \\
  \hline
\end{tabular}
\label{tab2}
\end{table}
\par
Now we move on to investigate the effect of temperature on the crossover time $\tau_s$ that marks the switch over to a completely reaction dominant dynamics. As already mentioned, $\tau_s$ is calculated using the prescription embedded in Eq.\ \eqref{eta_eqn} with $\eta=1$. In Fig.\ \ref{taus_diffT_pr}(a) we show the plots of the $\tau_s$ as a function of $p_r$, at different $T$. All the data seem to be parallel to each other except for the data points at $p_r=10^{-1}$ for $T=0.7T_c$ and $0.8T_c$. 
Possibly, the crossover is not so sharp at high $T$, as rate of both phase segregation and  \emph{interconversion} reaction get enhanced due to huge thermal fluctuations. Hence, the criterion used in Eq.\ \eqref{eta_eqn} fails to  unambiguously extract the crossover time $\tau_s$. Nonetheless, we obtain reasonable results when we fit the ansatz in Eq.\ \eqref{taus_fit} to the data at different $T$, the results of which are tabulated in Table\ \ref{tab2}. From the values quoted there we calculate the average prefactor $A=17.97\pm 3.1$ and $x=1.05\pm0.07$, which of course, have significant error bars. Nevertheless, in Fig.\ \ref{taus_diffT_pr}(a) we plot a dashed line representing Eq.\ \eqref{taus_fit} with the above quoted values. The data indeed look more or less parallel to the dashed line except for the points representing $p_r=10^{-4}$ at $T=0.7T_c$ and $0.8T_c$. 
\par
The variation of $\tau_s$ with $T$ for fixed values of the reaction probability $p_r$ are shown in Fig.\ \ref{taus_diffT_pr}(b). The almost flat behavior of the data for all cases indicates that the crossover is very weakly dependent on the temperature, although for purely segregating system it has been shown that the relaxation time shows an Arrhenius behavior \cite{Thwal1}. This of course is the courtesy of the effect of presence of \emph{interconversion} reaction even when the dynamics is dominated by segregation.

\section{Conclusion}\label{conclusion}
To summarize, we have presented results on the effects of interplay of phase segregation and \emph{interconversion} reaction on the kinetics of domain growth. For that purpose we have performed Metropolis MC simulations of the nearest neighbor Ising model in a square lattice, governed by both conserved and nonconserved dynamics. The system mimics a phase-segregating \emph{isomeric} binary mixture. Starting from such a homogeneous binary mixture, we have studied the nonequilibrium kinetics when the system is quenched below the demixing or critical temperature. Due to the presence of the reaction, in the asymptotic limit one of the \emph{isomers} will emerge as the \emph{winner}, i.e., it  will be present as the majority, and the other \emph{isomer} will perish which we refer to as the \emph{loser}. Instead of monitoring the average domain size of both \emph{isomers}, we have studied the kinetics of both of them separately.
\par
Our results show that for higher reaction probability $p_r$, the dynamics of the system resembles ferromagnetic ordering throughout the evolution. There one observes the usual scaling of the two-point equal-time correlation function $C(r,t)$ and the structure factor $S(k,t)$. As $p_r$ decreases, these quantities show signature of the strong interplay between the segregation dynamics and the reaction dynamics, thereby, the overall scaling picture does not hold anymore. For very low $p_r$, instead one observes an early-time scaling that resembles the conserved dynamics due to phase segregation. However, the universal Porod-tail behavior of $S(k,t) \sim k^{-3}$ remains unaffected for any choice of $p_r$.
\par
While for large $p_r$ the time dependence of the average domain size of the \emph{winner} $\ell_1(t)$ and \emph{loser} $\ell_2(t)$ show behavior consistent with ferromagnetic ordering, with decrease of $p_r$ the interplay of the conserved and noconserved dynamics show up. In fact, one observes a crossover in the growth from an initial slow $\ell_1(t)\sim t^{1/7}$ behavior to the usual $\ell_1(t) \sim t^{1/2}$ growth at late times. The time dependence of $\ell_2(t)$ for small $p_r$ also show a similar crossover with initially growing similarly with the power law $\ell_2(t) \sim t^{1/7}$. Afterwards, of course, $\ell_2(t)$ grows slower than $\ell_2(t)\sim t^{1/2}$. At very late time $\ell_2(t)$ starts to decay and asymptotically almost vanishes. We caution the reader that a proper theoretical understanding of the numerically observed early-time slow growth $\sim t^{1/7}$ is still required.

\par
The observation of the crossover in the growth prompted us to also calculate a crossover time $\tau_s$ from the time dependence of the difference between average domain size of the \emph{winner} and \emph{loser}, which at late times grows as $\ell_1-\ell_2 \sim t$. Our data for $\tau_s$ as a function of $p_r$ appears to follow the power-law scaling $\tau_s \sim p_r^{-x}$, where we obtained $x=1.05$, almost independent of temperature. The crossover time $\tau_s$ also appears to be  independent of temperature for fixed values of $p_r$.
\par
A facile adaptation of this work would be to consider a more complex reaction involving more than two species, for which MC simulation of an analogous $q$-state Potts model seems to be an automatic choice \cite{Potts_RMP,majumder2018Potts,Janke_CP}.
In future, it would also be interesting to study the effect of such interplay of phase segregation and \emph{interconversion} reaction on the other aspect of nonequilibrium dynamics of phase transition, i.e., aging and related scaling \cite{henkel_book,midya2014,midya2015}. Given the recent growing interest in nonequilibrium dynamics of long-range systems, it would certainly be worth to investigate this interplay of conserved and nonconserved dynamics using the long range Ising model with power-law interaction \cite{christiansen2019,christiansen2020,christiansen2021,mueller2022}. Lastly, to invoke more real effects of modeling such a system it would be challenging to construct a similar model where the reaction is happening in solutions. For that one needs to perform MD simulations of a fluid-like system where the role of hydrodynamics has to be taken into account \cite{majumder2011EPL,MajumderPRL}. 
\acknowledgments
The work was funded by the Science and Engineering Research Board (SERB), Govt.\ of India in the form of a Ramanujan Fellowship (file no.\ RJF/2021/000044).


\begin{thebibliography}{56}%
\makeatletter
\providecommand \@ifxundefined [1]{%
 \@ifx{#1\undefined}
}%
\providecommand \@ifnum [1]{%
 \ifnum #1\expandafter \@firstoftwo
 \else \expandafter \@secondoftwo
 \fi
}%
\providecommand \@ifx [1]{%
 \ifx #1\expandafter \@firstoftwo
 \else \expandafter \@secondoftwo
 \fi
}%
\providecommand \natexlab [1]{#1}%
\providecommand \enquote  [1]{``#1''}%
\providecommand \bibnamefont  [1]{#1}%
\providecommand \bibfnamefont [1]{#1}%
\providecommand \citenamefont [1]{#1}%
\providecommand \href@noop [0]{\@secondoftwo}%
\providecommand \href [0]{\begingroup \@sanitize@url \@href}%
\providecommand \@href[1]{\@@startlink{#1}\@@href}%
\providecommand \@@href[1]{\endgroup#1\@@endlink}%
\providecommand \@sanitize@url [0]{\catcode `\\12\catcode `\$12\catcode
  `\&12\catcode `\#12\catcode `\^12\catcode `\_12\catcode `\%12\relax}%
\providecommand \@@startlink[1]{}%
\providecommand \@@endlink[0]{}%
\providecommand \url  [0]{\begingroup\@sanitize@url \@url }%
\providecommand \@url [1]{\endgroup\@href {#1}{\urlprefix }}%
\providecommand \urlprefix  [0]{URL }%
\providecommand \Eprint [0]{\href }%
\providecommand \doibase [0]{http://dx.doi.org/}%
\providecommand \selectlanguage [0]{\@gobble}%
\providecommand \bibinfo  [0]{\@secondoftwo}%
\providecommand \bibfield  [0]{\@secondoftwo}%
\providecommand \translation [1]{[#1]}%
\providecommand \BibitemOpen [0]{}%
\providecommand \bibitemStop [0]{}%
\providecommand \bibitemNoStop [0]{.\EOS\space}%
\providecommand \EOS [0]{\spacefactor3000\relax}%
\providecommand \BibitemShut  [1]{\csname bibitem#1\endcsname}%
\let\auto@bib@innerbib\@empty
%</preamble>
\bibitem [{\citenamefont {Bray}(2002)}]{bray2002}%
  \BibitemOpen
  \bibfield  {author} {\bibinfo {author} {\bibfnamefont {A.J.}\ \bibnamefont
  {Bray}},\ }\bibfield  {title} {\enquote {\bibinfo {title} {Theory of
  phase-ordering kinetics},}\ }\href@noop {} {\bibfield  {journal} {\bibinfo
  {journal} {Adv. Phys.}\ }\textbf {\bibinfo {volume} {51}},\ \bibinfo {pages}
  {481--587} (\bibinfo {year} {2002})}\BibitemShut {NoStop}%
\bibitem [{\citenamefont {Puri}\ and\ \citenamefont
  {Wadhawan}(2009)}]{puri_book}%
  \BibitemOpen
  \bibinfo {editor} {\bibfnamefont {S.}~\bibnamefont {Puri}}\ and\ \bibinfo
  {editor} {\bibfnamefont {V.}~\bibnamefont {Wadhawan}},\ eds.,\ \href@noop {}
  {\emph {\bibinfo {title} {{K}inetics of {P}hase {T}ransitions}}}\ (\bibinfo
  {publisher} {CRC Press, Boca Raton},\ \bibinfo {year} {2009})\BibitemShut
  {NoStop}%
\bibitem [{\citenamefont {Stanley}(1971)}]{stanley1971}%
  \BibitemOpen
  \bibfield  {author} {\bibinfo {author} {\bibfnamefont {H.E.}\ \bibnamefont
  {Stanley}},\ }\href@noop {} {\emph {\bibinfo {title} {Introduction to phase
  transitions and critical phenomena}}}\ (\bibinfo  {publisher} {Clarendon
  Press, Oxford},\ \bibinfo {year} {1971})\BibitemShut {NoStop}%
\bibitem [{\citenamefont {Fisher}(1974)}]{fisher1974}%
  \BibitemOpen
  \bibfield  {author} {\bibinfo {author} {\bibfnamefont {M.E.}\ \bibnamefont
  {Fisher}},\ }\bibfield  {title} {\enquote {\bibinfo {title} {The
  renormalization group in the theory of critical behavior},}\ }\href@noop {}
  {\bibfield  {journal} {\bibinfo  {journal} {Rev. Mod. Phys.}\ }\textbf
  {\bibinfo {volume} {46}},\ \bibinfo {pages} {597} (\bibinfo {year}
  {1974})}\BibitemShut {NoStop}%
\bibitem [{\citenamefont {Hohenberg}\ and\ \citenamefont
  {Halperin}(1977)}]{Hohenberg1977}%
  \BibitemOpen
  \bibfield  {author} {\bibinfo {author} {\bibfnamefont {P.C.}\ \bibnamefont
  {Hohenberg}}\ and\ \bibinfo {author} {\bibfnamefont {B.I.}\ \bibnamefont
  {Halperin}},\ }\bibfield  {title} {\enquote {\bibinfo {title} {Theory of
  dynamic critical phenomena},}\ }\href@noop {} {\bibfield  {journal} {\bibinfo
   {journal} {Rev. Mod. Phys.}\ }\textbf {\bibinfo {volume} {49}},\ \bibinfo
  {pages} {435--479} (\bibinfo {year} {1977})}\BibitemShut {NoStop}%
\bibitem [{\citenamefont {Domb}\ \emph {et~al.}(1972-2001)\citenamefont {Domb},
  \citenamefont {Green},\ and\ \citenamefont {Lebowtiz}}]{domb2000}%
  \BibitemOpen
  \bibinfo {editor} {\bibfnamefont {C.}~\bibnamefont {Domb}}, \bibinfo {editor}
  {\bibfnamefont {M.S.}\ \bibnamefont {Green}}, \ and\ \bibinfo {editor}
  {\bibfnamefont {J.L.}\ \bibnamefont {Lebowtiz}},\ eds.,\ \href@noop {} {\emph
  {\bibinfo {title} {Phase transitions and critical phenomena}}},\ Vol.\
  \bibinfo {volume} {1-20}\ (\bibinfo  {publisher} {Elsevier},\ \bibinfo {year}
  {1972-2001})\BibitemShut {NoStop}%
\bibitem [{\citenamefont {Siggia}(1979)}]{Siggia1979}%
  \BibitemOpen
  \bibfield  {author} {\bibinfo {author} {\bibfnamefont {E.D.}\ \bibnamefont
  {Siggia}},\ }\bibfield  {title} {\enquote {\bibinfo {title} {Late stages of
  spinodal decomposition in binary mixtures},}\ }\href@noop {} {\bibfield
  {journal} {\bibinfo  {journal} {Phys. Rev. A}\ }\textbf {\bibinfo {volume}
  {20}},\ \bibinfo {pages} {595--605} (\bibinfo {year} {1979})}\BibitemShut
  {NoStop}%
\bibitem [{\citenamefont {Furukawa}(1985)}]{Furukawa1985}%
  \BibitemOpen
  \bibfield  {author} {\bibinfo {author} {\bibfnamefont {H.}~\bibnamefont
  {Furukawa}},\ }\bibfield  {title} {\enquote {\bibinfo {title} {Effect of
  inertia on droplet growth in a fluid},}\ }\href@noop {} {\bibfield  {journal}
  {\bibinfo  {journal} {Phys. Rev. A}\ }\textbf {\bibinfo {volume} {31}},\
  \bibinfo {pages} {1103--1108} (\bibinfo {year} {1985})}\BibitemShut {NoStop}%
\bibitem [{\citenamefont {Furukawa}(1987)}]{Furukawa1987}%
  \BibitemOpen
  \bibfield  {author} {\bibinfo {author} {\bibfnamefont {H.}~\bibnamefont
  {Furukawa}},\ }\bibfield  {title} {\enquote {\bibinfo {title} {Turbulent
  growth of percolated droplets in phase-separating fluids},}\ }\href@noop {}
  {\bibfield  {journal} {\bibinfo  {journal} {Phys. Rev. A}\ }\textbf {\bibinfo
  {volume} {36}},\ \bibinfo {pages} {2288--2292} (\bibinfo {year}
  {1987})}\BibitemShut {NoStop}%
\bibitem [{\citenamefont {Allen}\ and\ \citenamefont {Cahn}(1979)}]{allen1979}%
  \BibitemOpen
  \bibfield  {author} {\bibinfo {author} {\bibfnamefont {S.M.}\ \bibnamefont
  {Allen}}\ and\ \bibinfo {author} {\bibfnamefont {J.W.}\ \bibnamefont
  {Cahn}},\ }\bibfield  {title} {\enquote {\bibinfo {title} {A microscopic
  theory for antiphase boundary motion and its application to antiphase domain
  coarsening},}\ }\href@noop {} {\bibfield  {journal} {\bibinfo  {journal}
  {Acta Metall.}\ }\textbf {\bibinfo {volume} {27}},\ \bibinfo {pages}
  {1085--1095} (\bibinfo {year} {1979})}\BibitemShut {NoStop}%
\bibitem [{\citenamefont {Lifshitz}\ and\ \citenamefont
  {Slyozov}(1961)}]{lifshitz1961}%
  \BibitemOpen
  \bibfield  {author} {\bibinfo {author} {\bibfnamefont {I.M.}\ \bibnamefont
  {Lifshitz}}\ and\ \bibinfo {author} {\bibfnamefont {V.V.}\ \bibnamefont
  {Slyozov}},\ }\bibfield  {title} {\enquote {\bibinfo {title} {The kinetics of
  precipitation from supersaturated solid solutions},}\ }\href@noop {}
  {\bibfield  {journal} {\bibinfo  {journal} {J. Phys. Chem. Solids}\ }\textbf
  {\bibinfo {volume} {19}},\ \bibinfo {pages} {35--50} (\bibinfo {year}
  {1961})}\BibitemShut {NoStop}%
\bibitem [{\citenamefont {Lifshitz}(1962)}]{lifshitz1962}%
  \BibitemOpen
  \bibfield  {author} {\bibinfo {author} {\bibfnamefont {I.M.}\ \bibnamefont
  {Lifshitz}},\ }\bibfield  {title} {\enquote {\bibinfo {title} {Kinetics of
  ordering during second-order phase transitions},}\ }\href@noop {} {\bibfield
  {journal} {\bibinfo  {journal} {Sov. Phys. JETP}\ }\textbf {\bibinfo {volume}
  {15}},\ \bibinfo {pages} {939} (\bibinfo {year} {1962})}\BibitemShut
  {NoStop}%
\bibitem [{\citenamefont {Marko}\ and\ \citenamefont
  {Barkema}(1995)}]{marko1995}%
  \BibitemOpen
  \bibfield  {author} {\bibinfo {author} {\bibfnamefont {J.F.}\ \bibnamefont
  {Marko}}\ and\ \bibinfo {author} {\bibfnamefont {G.T.}\ \bibnamefont
  {Barkema}},\ }\bibfield  {title} {\enquote {\bibinfo {title} {Phase ordering
  in the {I}sing model with conserved spin},}\ }\href@noop {} {\bibfield
  {journal} {\bibinfo  {journal} {Phys. Rev. E}\ }\textbf {\bibinfo {volume}
  {52}},\ \bibinfo {pages} {2522} (\bibinfo {year} {1995})}\BibitemShut
  {NoStop}%
\bibitem [{\citenamefont {Amar}\ \emph {et~al.}(1988)\citenamefont {Amar},
  \citenamefont {Sullivan},\ and\ \citenamefont {Mountain}}]{amar1988}%
  \BibitemOpen
  \bibfield  {author} {\bibinfo {author} {\bibfnamefont {J.G.}\ \bibnamefont
  {Amar}}, \bibinfo {author} {\bibfnamefont {F.E.}\ \bibnamefont {Sullivan}}, \
  and\ \bibinfo {author} {\bibfnamefont {R.D.}\ \bibnamefont {Mountain}},\
  }\bibfield  {title} {\enquote {\bibinfo {title} {{M}onte {C}arlo study of
  growth in the two-dimensional spin-exchange kinetic {I}sing model},}\
  }\href@noop {} {\bibfield  {journal} {\bibinfo  {journal} {Phys. Rev. B}\
  }\textbf {\bibinfo {volume} {37}},\ \bibinfo {pages} {196} (\bibinfo {year}
  {1988})}\BibitemShut {NoStop}%
\bibitem [{\citenamefont {Majumder}\ and\ \citenamefont
  {Das}(2010)}]{majumder2010}%
  \BibitemOpen
  \bibfield  {author} {\bibinfo {author} {\bibfnamefont {S.}~\bibnamefont
  {Majumder}}\ and\ \bibinfo {author} {\bibfnamefont {S.K.}\ \bibnamefont
  {Das}},\ }\bibfield  {title} {\enquote {\bibinfo {title} {Domain coarsening
  in two dimensions: {C}onserved dynamics and finite-size scaling},}\
  }\href@noop {} {\bibfield  {journal} {\bibinfo  {journal} {Phys. Rev. E}\
  }\textbf {\bibinfo {volume} {81}},\ \bibinfo {pages} {050102} (\bibinfo
  {year} {2010})}\BibitemShut {NoStop}%
\bibitem [{\citenamefont {Majumder}\ and\ \citenamefont
  {Das}(2011{\natexlab{a}})}]{majumder2011}%
  \BibitemOpen
  \bibfield  {author} {\bibinfo {author} {\bibfnamefont {S.}~\bibnamefont
  {Majumder}}\ and\ \bibinfo {author} {\bibfnamefont {S.K.}\ \bibnamefont
  {Das}},\ }\bibfield  {title} {\enquote {\bibinfo {title} {Diffusive domain
  coarsening: Early time dynamics and finite-size effects},}\ }\href@noop {}
  {\bibfield  {journal} {\bibinfo  {journal} {Phys. Rev. E}\ }\textbf {\bibinfo
  {volume} {84}},\ \bibinfo {pages} {021110} (\bibinfo {year}
  {2011}{\natexlab{a}})}\BibitemShut {NoStop}%
\bibitem [{\citenamefont {Majumder}\ and\ \citenamefont
  {Das}(2013{\natexlab{a}})}]{majumder2013}%
  \BibitemOpen
  \bibfield  {author} {\bibinfo {author} {\bibfnamefont {S.}~\bibnamefont
  {Majumder}}\ and\ \bibinfo {author} {\bibfnamefont {S.K.}\ \bibnamefont
  {Das}},\ }\bibfield  {title} {\enquote {\bibinfo {title} {Temperature and
  composition dependence of kinetics of phase separation in solid binary
  mixtures},}\ }\href@noop {} {\bibfield  {journal} {\bibinfo  {journal} {Phys.
  Chem. Chem. Phys.}\ }\textbf {\bibinfo {volume} {15}},\ \bibinfo {pages}
  {13209--13218} (\bibinfo {year} {2013}{\natexlab{a}})}\BibitemShut {NoStop}%
\bibitem [{\citenamefont {Christiansen}\ \emph {et~al.}(2019)\citenamefont
  {Christiansen}, \citenamefont {Majumder},\ and\ \citenamefont
  {Janke}}]{christiansen2019}%
  \BibitemOpen
  \bibfield  {author} {\bibinfo {author} {\bibfnamefont {Henrik}\ \bibnamefont
  {Christiansen}}, \bibinfo {author} {\bibfnamefont {Suman}\ \bibnamefont
  {Majumder}}, \ and\ \bibinfo {author} {\bibfnamefont {Wolfhard}\ \bibnamefont
  {Janke}},\ }\bibfield  {title} {\enquote {\bibinfo {title} {Phase ordering
  kinetics of the long-range {I}sing model},}\ }\href@noop {} {\bibfield
  {journal} {\bibinfo  {journal} {Phys. Rev. E}\ }\textbf {\bibinfo {volume}
  {99}},\ \bibinfo {pages} {011301} (\bibinfo {year} {2019})}\BibitemShut
  {NoStop}%
\bibitem [{\citenamefont {Christiansen}\ \emph {et~al.}(2020)\citenamefont
  {Christiansen}, \citenamefont {Majumder}, \citenamefont {Henkel},\ and\
  \citenamefont {Janke}}]{christiansen2020}%
  \BibitemOpen
  \bibfield  {author} {\bibinfo {author} {\bibfnamefont {H.}~\bibnamefont
  {Christiansen}}, \bibinfo {author} {\bibfnamefont {S.}~\bibnamefont
  {Majumder}}, \bibinfo {author} {\bibfnamefont {M.}~\bibnamefont {Henkel}}, \
  and\ \bibinfo {author} {\bibfnamefont {W.}~\bibnamefont {Janke}},\ }\bibfield
   {title} {\enquote {\bibinfo {title} {Aging in the long-range {I}sing
  model},}\ }\href@noop {} {\bibfield  {journal} {\bibinfo  {journal} {Phys.
  Rev. Lett.}\ }\textbf {\bibinfo {volume} {125}},\ \bibinfo {pages} {180601}
  (\bibinfo {year} {2020})}\BibitemShut {NoStop}%
\bibitem [{\citenamefont {Christiansen}\ \emph {et~al.}(2021)\citenamefont
  {Christiansen}, \citenamefont {Majumder},\ and\ \citenamefont
  {Janke}}]{christiansen2021}%
  \BibitemOpen
  \bibfield  {author} {\bibinfo {author} {\bibfnamefont {H.}~\bibnamefont
  {Christiansen}}, \bibinfo {author} {\bibfnamefont {S.}~\bibnamefont
  {Majumder}}, \ and\ \bibinfo {author} {\bibfnamefont {W.}~\bibnamefont
  {Janke}},\ }\bibfield  {title} {\enquote {\bibinfo {title} {Zero-temperature
  coarsening in the two-dimensional long-range {I}sing model},}\ }\href@noop {}
  {\bibfield  {journal} {\bibinfo  {journal} {Phys. Rev. E}\ }\textbf {\bibinfo
  {volume} {103}},\ \bibinfo {pages} {052122} (\bibinfo {year}
  {2021})}\BibitemShut {NoStop}%
\bibitem [{\citenamefont {M{\"u}ller}\ \emph {et~al.}(2022)\citenamefont
  {M{\"u}ller}, \citenamefont {Christiansen},\ and\ \citenamefont
  {Janke}}]{mueller2022}%
  \BibitemOpen
  \bibfield  {author} {\bibinfo {author} {\bibfnamefont {F.}~\bibnamefont
  {M{\"u}ller}}, \bibinfo {author} {\bibfnamefont {H.}~\bibnamefont
  {Christiansen}}, \ and\ \bibinfo {author} {\bibfnamefont {W.}~\bibnamefont
  {Janke}},\ }\bibfield  {title} {\enquote {\bibinfo {title} {Phase-separation
  kinetics in the two-dimensional long-range {I}sing model},}\ }\href@noop {}
  {\bibfield  {journal} {\bibinfo  {journal} {Phys. Rev. Lett.}\ }\textbf
  {\bibinfo {volume} {129}},\ \bibinfo {pages} {240601} (\bibinfo {year}
  {2022})}\BibitemShut {NoStop}%
\bibitem [{\citenamefont {McNaught}\ \emph {et~al.}(1997)\citenamefont
  {McNaught}, \citenamefont {Wilkinson} \emph {et~al.}}]{mcnaught_book}%
  \BibitemOpen
  \bibfield  {author} {\bibinfo {author} {\bibfnamefont {A.D.}\ \bibnamefont
  {McNaught}}, \bibinfo {author} {\bibfnamefont {A}~\bibnamefont {Wilkinson}},
  \emph {et~al.},\ }\href@noop {} {\emph {\bibinfo {title} {{C}ompendium of
  {C}hemical {T}erminology}}},\ Vol.\ \bibinfo {volume} {1669}\ (\bibinfo
  {publisher} {Blackwell Science Oxford},\ \bibinfo {year} {1997})\BibitemShut
  {NoStop}%
\bibitem [{\citenamefont {Soai}\ \emph {et~al.}(1995)\citenamefont {Soai},
  \citenamefont {Shibata}, \citenamefont {Morioka},\ and\ \citenamefont
  {Choji}}]{soai1995}%
  \BibitemOpen
  \bibfield  {author} {\bibinfo {author} {\bibfnamefont {K.}~\bibnamefont
  {Soai}}, \bibinfo {author} {\bibfnamefont {T.}~\bibnamefont {Shibata}},
  \bibinfo {author} {\bibfnamefont {H.}~\bibnamefont {Morioka}}, \ and\
  \bibinfo {author} {\bibfnamefont {K.}~\bibnamefont {Choji}},\ }\bibfield
  {title} {\enquote {\bibinfo {title} {Asymmetric autocatalysis and
  amplification of enantiomeric excess of a chiral molecule},}\ }\href@noop {}
  {\bibfield  {journal} {\bibinfo  {journal} {Nature}\ }\textbf {\bibinfo
  {volume} {378}},\ \bibinfo {pages} {767--768} (\bibinfo {year}
  {1995})}\BibitemShut {NoStop}%
\bibitem [{\citenamefont {Shibata}\ \emph
  {et~al.}(1996{\natexlab{a}})\citenamefont {Shibata}, \citenamefont {Choji},
  \citenamefont {Hayase}, \citenamefont {Aizu},\ and\ \citenamefont
  {Soai}}]{shibata1996}%
  \BibitemOpen
  \bibfield  {author} {\bibinfo {author} {\bibfnamefont {T.}~\bibnamefont
  {Shibata}}, \bibinfo {author} {\bibfnamefont {K.}~\bibnamefont {Choji}},
  \bibinfo {author} {\bibfnamefont {T.}~\bibnamefont {Hayase}}, \bibinfo
  {author} {\bibfnamefont {Y.}~\bibnamefont {Aizu}}, \ and\ \bibinfo {author}
  {\bibfnamefont {K.}~\bibnamefont {Soai}},\ }\bibfield  {title} {\enquote
  {\bibinfo {title} {Asymmetric autocatalytic reaction of 3-quinolylalkanol
  with amplification of enantiomeric excess},}\ }\href@noop {} {\bibfield
  {journal} {\bibinfo  {journal} {Chem. Comm.}\ }\textbf {\bibinfo {volume}
  {10}},\ \bibinfo {pages} {1235--1236} (\bibinfo {year}
  {1996}{\natexlab{a}})}\BibitemShut {NoStop}%
\bibitem [{\citenamefont {Shibata}\ \emph
  {et~al.}(1996{\natexlab{b}})\citenamefont {Shibata}, \citenamefont {Morioka},
  \citenamefont {Hayase}, \citenamefont {Choji},\ and\ \citenamefont
  {Soai}}]{shibata1996_jacs}%
  \BibitemOpen
  \bibfield  {author} {\bibinfo {author} {\bibfnamefont {T.}~\bibnamefont
  {Shibata}}, \bibinfo {author} {\bibfnamefont {H.}~\bibnamefont {Morioka}},
  \bibinfo {author} {\bibfnamefont {T.}~\bibnamefont {Hayase}}, \bibinfo
  {author} {\bibfnamefont {K.}~\bibnamefont {Choji}}, \ and\ \bibinfo {author}
  {\bibfnamefont {K.}~\bibnamefont {Soai}},\ }\bibfield  {title} {\enquote
  {\bibinfo {title} {Highly enantioselective catalytic asymmetric
  automultiplication of chiral pyrimidyl alcohol},}\ }\href@noop {} {\bibfield
  {journal} {\bibinfo  {journal} {J. Am. Chem. Soc.}\ }\textbf {\bibinfo
  {volume} {118}},\ \bibinfo {pages} {471--472} (\bibinfo {year}
  {1996}{\natexlab{b}})}\BibitemShut {NoStop}%
\bibitem [{\citenamefont {Tran-Cong}\ and\ \citenamefont
  {Harada}(1996)}]{tran1996reaction}%
  \BibitemOpen
  \bibfield  {author} {\bibinfo {author} {\bibfnamefont {Q.}~\bibnamefont
  {Tran-Cong}}\ and\ \bibinfo {author} {\bibfnamefont {A.}~\bibnamefont
  {Harada}},\ }\bibfield  {title} {\enquote {\bibinfo {title} {Reaction-induced
  ordering phenomena in binary polymer mixtures},}\ }\href@noop {} {\bibfield
  {journal} {\bibinfo  {journal} {Phys. Rev. Lett.}\ }\textbf {\bibinfo
  {volume} {76}},\ \bibinfo {pages} {1162--1165} (\bibinfo {year}
  {1996})}\BibitemShut {NoStop}%
\bibitem [{\citenamefont {Tran-Cong}\ \emph {et~al.}(1997)\citenamefont
  {Tran-Cong}, \citenamefont {Ohta},\ and\ \citenamefont {Urakawa}}]{Qui1997}%
  \BibitemOpen
  \bibfield  {author} {\bibinfo {author} {\bibfnamefont {Q.}~\bibnamefont
  {Tran-Cong}}, \bibinfo {author} {\bibfnamefont {T.}~\bibnamefont {Ohta}}, \
  and\ \bibinfo {author} {\bibfnamefont {O.}~\bibnamefont {Urakawa}},\
  }\bibfield  {title} {\enquote {\bibinfo {title} {Soft-mode suppression in the
  phase separation of binary polymer mixtures driven by a reversible chemical
  reaction},}\ }\href@noop {} {\bibfield  {journal} {\bibinfo  {journal} {Phys.
  Rev. E}\ }\textbf {\bibinfo {volume} {56}},\ \bibinfo {pages} {R59--R62}
  (\bibinfo {year} {1997})}\BibitemShut {NoStop}%
\bibitem [{\citenamefont {Ohta}\ \emph {et~al.}(1998)\citenamefont {Ohta},
  \citenamefont {Urakawa},\ and\ \citenamefont {Tran-Cong}}]{ohta1998}%
  \BibitemOpen
  \bibfield  {author} {\bibinfo {author} {\bibfnamefont {T.}~\bibnamefont
  {Ohta}}, \bibinfo {author} {\bibfnamefont {O.}~\bibnamefont {Urakawa}}, \
  and\ \bibinfo {author} {\bibfnamefont {Q.}~\bibnamefont {Tran-Cong}},\
  }\bibfield  {title} {\enquote {\bibinfo {title} {Phase separation of binary
  polymer blends driven by photoisomerization: {A}n example for a
  wavelength-selection process in polymers},}\ }\href@noop {} {\bibfield
  {journal} {\bibinfo  {journal} {Macromolecules}\ }\textbf {\bibinfo {volume}
  {31}},\ \bibinfo {pages} {6845--6854} (\bibinfo {year} {1998})}\BibitemShut
  {NoStop}%
\bibitem [{\citenamefont {Viedma}(2005)}]{viedma2005}%
  \BibitemOpen
  \bibfield  {author} {\bibinfo {author} {\bibfnamefont {C.}~\bibnamefont
  {Viedma}},\ }\bibfield  {title} {\enquote {\bibinfo {title} {Chiral symmetry
  breaking during crystallization: complete chiral purity induced by nonlinear
  autocatalysis and recycling},}\ }\href@noop {} {\bibfield  {journal}
  {\bibinfo  {journal} {Phys. Rev. Lett.}\ }\textbf {\bibinfo {volume} {94}},\
  \bibinfo {pages} {065504} (\bibinfo {year} {2005})}\BibitemShut {NoStop}%
\bibitem [{\citenamefont {Lombardo}\ \emph {et~al.}(2009)\citenamefont
  {Lombardo}, \citenamefont {Stillinger},\ and\ \citenamefont
  {Debenedetti}}]{lombardo2009}%
  \BibitemOpen
  \bibfield  {author} {\bibinfo {author} {\bibfnamefont {T.G.}\ \bibnamefont
  {Lombardo}}, \bibinfo {author} {\bibfnamefont {F.H.}\ \bibnamefont
  {Stillinger}}, \ and\ \bibinfo {author} {\bibfnamefont {P.G.}\ \bibnamefont
  {Debenedetti}},\ }\bibfield  {title} {\enquote {\bibinfo {title}
  {Thermodynamic mechanism for solution phase chiral amplification via a
  lattice model},}\ }\href@noop {} {\bibfield  {journal} {\bibinfo  {journal}
  {Proc. Nat. Acad. Sci.}\ }\textbf {\bibinfo {volume} {106}},\ \bibinfo
  {pages} {15131--15135} (\bibinfo {year} {2009})}\BibitemShut {NoStop}%
\bibitem [{\citenamefont {Glotzer}\ \emph {et~al.}(1994)\citenamefont
  {Glotzer}, \citenamefont {Stauffer},\ and\ \citenamefont
  {Jan}}]{glotzer1994}%
  \BibitemOpen
  \bibfield  {author} {\bibinfo {author} {\bibfnamefont {S.C.}\ \bibnamefont
  {Glotzer}}, \bibinfo {author} {\bibfnamefont {D.}~\bibnamefont {Stauffer}}, \
  and\ \bibinfo {author} {\bibfnamefont {N.}~\bibnamefont {Jan}},\ }\bibfield
  {title} {\enquote {\bibinfo {title} {Monte carlo simulations of phase
  separation in chemically reactive binary mixtures},}\ }\href@noop {}
  {\bibfield  {journal} {\bibinfo  {journal} {Phys. Rev. Lett.}\ }\textbf
  {\bibinfo {volume} {72}},\ \bibinfo {pages} {4109--4112} (\bibinfo {year}
  {1994})}\BibitemShut {NoStop}%
\bibitem [{\citenamefont {Puri}\ and\ \citenamefont {Frisch}(1994)}]{puri1994}%
  \BibitemOpen
  \bibfield  {author} {\bibinfo {author} {\bibfnamefont {S.}~\bibnamefont
  {Puri}}\ and\ \bibinfo {author} {\bibfnamefont {H.L.}\ \bibnamefont
  {Frisch}},\ }\bibfield  {title} {\enquote {\bibinfo {title} {Segregation
  dynamics of binary mixtures with simple chemical reactions},}\ }\href@noop {}
  {\bibfield  {journal} {\bibinfo  {journal} {J. Phys. A.}\ }\textbf {\bibinfo
  {volume} {27}},\ \bibinfo {pages} {6027--6038} (\bibinfo {year}
  {1994})}\BibitemShut {NoStop}%
\bibitem [{\citenamefont {Glotzer}\ \emph {et~al.}(1995)\citenamefont
  {Glotzer}, \citenamefont {Di~Marzio},\ and\ \citenamefont
  {Muthukumar}}]{glotzer1995}%
  \BibitemOpen
  \bibfield  {author} {\bibinfo {author} {\bibfnamefont {S.C.}\ \bibnamefont
  {Glotzer}}, \bibinfo {author} {\bibfnamefont {E.A.}\ \bibnamefont
  {Di~Marzio}}, \ and\ \bibinfo {author} {\bibfnamefont {M.}~\bibnamefont
  {Muthukumar}},\ }\bibfield  {title} {\enquote {\bibinfo {title}
  {Reaction-controlled morphology of phase-separating mixtures},}\ }\href@noop
  {} {\bibfield  {journal} {\bibinfo  {journal} {Phys. Rev. Lett.}\ }\textbf
  {\bibinfo {volume} {74}},\ \bibinfo {pages} {2034--2037} (\bibinfo {year}
  {1995})}\BibitemShut {NoStop}%
\bibitem [{\citenamefont {Singh}\ \emph {et~al.}(2012)\citenamefont {Singh},
  \citenamefont {Puri},\ and\ \citenamefont {Dasgupta}}]{singh2012}%
  \BibitemOpen
  \bibfield  {author} {\bibinfo {author} {\bibfnamefont {A.}~\bibnamefont
  {Singh}}, \bibinfo {author} {\bibfnamefont {S.}~\bibnamefont {Puri}}, \ and\
  \bibinfo {author} {\bibfnamefont {C.}~\bibnamefont {Dasgupta}},\ }\bibfield
  {title} {\enquote {\bibinfo {title} {Growth kinetics of nanoclusters in
  solution},}\ }\href@noop {} {\bibfield  {journal} {\bibinfo  {journal} {J.
  Phys. Chem. B}\ }\textbf {\bibinfo {volume} {116}},\ \bibinfo {pages}
  {4519--4523} (\bibinfo {year} {2012})}\BibitemShut {NoStop}%
\bibitem [{\citenamefont {Shumovskyi}\ \emph {et~al.}(2021)\citenamefont
  {Shumovskyi}, \citenamefont {Longo}, \citenamefont {Buldyrev},\ and\
  \citenamefont {Anisimov}}]{shumovskyi2021}%
  \BibitemOpen
  \bibfield  {author} {\bibinfo {author} {\bibfnamefont {N.A.}\ \bibnamefont
  {Shumovskyi}}, \bibinfo {author} {\bibfnamefont {T.J.}\ \bibnamefont
  {Longo}}, \bibinfo {author} {\bibfnamefont {S.V.}\ \bibnamefont {Buldyrev}},
  \ and\ \bibinfo {author} {\bibfnamefont {M.A.}\ \bibnamefont {Anisimov}},\
  }\bibfield  {title} {\enquote {\bibinfo {title} {Phase amplification in
  spinodal decomposition of immiscible fluids with interconversion of
  species},}\ }\href@noop {} {\bibfield  {journal} {\bibinfo  {journal} {Phys.
  Rev. E}\ }\textbf {\bibinfo {volume} {103}},\ \bibinfo {pages} {L060101}
  (\bibinfo {year} {2021})}\BibitemShut {NoStop}%
\bibitem [{\citenamefont {Longo}\ and\ \citenamefont
  {Anisimov}(2022)}]{longo2022}%
  \BibitemOpen
  \bibfield  {author} {\bibinfo {author} {\bibfnamefont {T.J.}\ \bibnamefont
  {Longo}}\ and\ \bibinfo {author} {\bibfnamefont {M.A.}\ \bibnamefont
  {Anisimov}},\ }\bibfield  {title} {\enquote {\bibinfo {title} {Phase
  transitions affected by natural and forceful molecular interconversion},}\
  }\href@noop {} {\bibfield  {journal} {\bibinfo  {journal} {J. Chem. Phys.}\
  }\textbf {\bibinfo {volume} {156}},\ \bibinfo {pages} {084502} (\bibinfo
  {year} {2022})}\BibitemShut {NoStop}%
\bibitem [{\citenamefont {Longo}\ \emph {et~al.}(2023)\citenamefont {Longo},
  \citenamefont {Shumovskyi}, \citenamefont {Uralcan}, \citenamefont
  {Buldyrev}, \citenamefont {Anisimov},\ and\ \citenamefont
  {Debenedetti}}]{longo2023}%
  \BibitemOpen
  \bibfield  {author} {\bibinfo {author} {\bibfnamefont {T.J.}\ \bibnamefont
  {Longo}}, \bibinfo {author} {\bibfnamefont {N.A.}\ \bibnamefont
  {Shumovskyi}}, \bibinfo {author} {\bibfnamefont {B.}~\bibnamefont {Uralcan}},
  \bibinfo {author} {\bibfnamefont {S.V.}\ \bibnamefont {Buldyrev}}, \bibinfo
  {author} {\bibfnamefont {M.A.}\ \bibnamefont {Anisimov}}, \ and\ \bibinfo
  {author} {\bibfnamefont {P.G.}\ \bibnamefont {Debenedetti}},\ }\bibfield
  {title} {\enquote {\bibinfo {title} {Formation of dissipative structures in
  microscopic models of mixtures with species interconversion},}\ }\href@noop
  {} {\bibfield  {journal} {\bibinfo  {journal} {Proc. Nat. Acad. Sci.}\
  }\textbf {\bibinfo {volume} {120}},\ \bibinfo {pages} {e2215012120} (\bibinfo
  {year} {2023})}\BibitemShut {NoStop}%
\bibitem [{\citenamefont {Thwal}\ and\ \citenamefont
  {Majumder}(2023)}]{Thwal1}%
  \BibitemOpen
  \bibfield  {author} {\bibinfo {author} {\bibfnamefont {S.}~\bibnamefont
  {Thwal}}\ and\ \bibinfo {author} {\bibfnamefont {S.}~\bibnamefont
  {Majumder}},\ }\bibfield  {title} {\enquote {\bibinfo {title} {Segregation
  disrupts the {A}rrhenius behavior of an isomerization reaction},}\
  }\href@noop {} {\bibfield  {journal} {\bibinfo  {journal} {arXiv preprint
  arXiv:2308.02258}\ } (\bibinfo {year} {2023})}\BibitemShut {NoStop}%
\bibitem [{\citenamefont {Majumder}(2023)}]{majumder2023}%
  \BibitemOpen
  \bibfield  {author} {\bibinfo {author} {\bibfnamefont {Suman}\ \bibnamefont
  {Majumder}},\ }\bibfield  {title} {\enquote {\bibinfo {title} {Disentangling
  growth and decay of domains during phase ordering},}\ }\href@noop {}
  {\bibfield  {journal} {\bibinfo  {journal} {Phys. Rev. E}\ }\textbf {\bibinfo
  {volume} {107}},\ \bibinfo {pages} {034130} (\bibinfo {year}
  {2023})}\BibitemShut {NoStop}%
\bibitem [{\citenamefont {Onsager}(1944)}]{onsager1944}%
  \BibitemOpen
  \bibfield  {author} {\bibinfo {author} {\bibfnamefont {L.}~\bibnamefont
  {Onsager}},\ }\bibfield  {title} {\enquote {\bibinfo {title} {{C}rystal
  statistics. {I}. {A} two-dimensional model with an order-disorder
  transition},}\ }\href@noop {} {\bibfield  {journal} {\bibinfo  {journal}
  {Phys. Rev.}\ }\textbf {\bibinfo {volume} {65}},\ \bibinfo {pages} {117--149}
  (\bibinfo {year} {1944})}\BibitemShut {NoStop}%
\bibitem [{\citenamefont {Glauber}(1963)}]{glauber1963}%
  \BibitemOpen
  \bibfield  {author} {\bibinfo {author} {\bibfnamefont {R.J.}\ \bibnamefont
  {Glauber}},\ }\bibfield  {title} {\enquote {\bibinfo {title} {Time-dependent
  statistics of the {I}sing model},}\ }\href@noop {} {\bibfield  {journal}
  {\bibinfo  {journal} {J. Math. Phys.}\ }\textbf {\bibinfo {volume} {4}},\
  \bibinfo {pages} {294--307} (\bibinfo {year} {1963})}\BibitemShut {NoStop}%
\bibitem [{\citenamefont {Newman}\ and\ \citenamefont
  {Barkema}(1999)}]{barkema_book}%
  \BibitemOpen
  \bibfield  {author} {\bibinfo {author} {\bibfnamefont {M.E.J.}\ \bibnamefont
  {Newman}}\ and\ \bibinfo {author} {\bibfnamefont {G.T.}\ \bibnamefont
  {Barkema}},\ }\href@noop {} {\emph {\bibinfo {title} {Monte {C}arlo {M}ethods
  in {S}tatistical {P}hysics}}}\ (\bibinfo  {publisher} {Oxford University
  Press, Oxford},\ \bibinfo {year} {1999})\BibitemShut {NoStop}%
\bibitem [{\citenamefont {Landau}\ and\ \citenamefont
  {Binder}(2021)}]{landau_book}%
  \BibitemOpen
  \bibfield  {author} {\bibinfo {author} {\bibfnamefont {D.}~\bibnamefont
  {Landau}}\ and\ \bibinfo {author} {\bibfnamefont {K.}~\bibnamefont
  {Binder}},\ }\href@noop {} {\emph {\bibinfo {title} {A {G}uide to {M}onte
  {C}arlo {S}imulations in {S}tatistical {P}hysics}}}\ (\bibinfo  {publisher}
  {Cambridge University Press, Cambridge},\ \bibinfo {year} {2021})\BibitemShut
  {NoStop}%
\bibitem [{\citenamefont {Kawasaki}(1966)}]{kawasaki1966}%
  \BibitemOpen
  \bibfield  {author} {\bibinfo {author} {\bibfnamefont {K.}~\bibnamefont
  {Kawasaki}},\ }\bibfield  {title} {\enquote {\bibinfo {title} {Diffusion
  constants near the critical point for time-dependent {I}sing models.\ {I}},}\
  }\href@noop {} {\bibfield  {journal} {\bibinfo  {journal} {Phys. Rev.}\
  }\textbf {\bibinfo {volume} {145}},\ \bibinfo {pages} {224--230} (\bibinfo
  {year} {1966})}\BibitemShut {NoStop}%
\bibitem [{\citenamefont {Porod}(1982)}]{porod1982}%
  \BibitemOpen
  \bibfield  {author} {\bibinfo {author} {\bibfnamefont {G.}~\bibnamefont
  {Porod}},\ }\bibfield  {title} {\enquote {\bibinfo {title} {General
  theory},}\ }in\ \href@noop {} {\emph {\bibinfo {booktitle} {Small {A}ngle
  {X}-ray {S}cattering}}},\ \bibinfo {editor} {edited by\ \bibinfo {editor}
  {\bibfnamefont {O.}~\bibnamefont {Glatter}}\ and\ \bibinfo {editor}
  {\bibfnamefont {O.}~\bibnamefont {Kratky}}}\ (\bibinfo  {publisher} {Aacdemic
  Press, London},\ \bibinfo {year} {1982})\ pp.\ \bibinfo {pages}
  {17--35}\BibitemShut {NoStop}%
\bibitem [{\citenamefont {Olejarz}\ \emph {et~al.}(2013)\citenamefont
  {Olejarz}, \citenamefont {Krapivsky},\ and\ \citenamefont
  {Redner}}]{olejarz2013}%
  \BibitemOpen
  \bibfield  {author} {\bibinfo {author} {\bibfnamefont {J.}~\bibnamefont
  {Olejarz}}, \bibinfo {author} {\bibfnamefont {P.L.}\ \bibnamefont
  {Krapivsky}}, \ and\ \bibinfo {author} {\bibfnamefont {S.}~\bibnamefont
  {Redner}},\ }\bibfield  {title} {\enquote {\bibinfo {title} {Zero-temperature
  coarsening in the 2d {P}otts model},}\ }\href@noop {} {\bibfield  {journal}
  {\bibinfo  {journal} {J. Stat. Mech.: Theo. Exp.}\ }\textbf {\bibinfo
  {volume} {2013}},\ \bibinfo {pages} {P06018} (\bibinfo {year}
  {2013})}\BibitemShut {NoStop}%
\bibitem [{\citenamefont {Efron}(1982)}]{efron1982}%
  \BibitemOpen
  \bibfield  {author} {\bibinfo {author} {\bibfnamefont {B.}~\bibnamefont
  {Efron}},\ }\href@noop {} {\emph {\bibinfo {title} {The {J}ackknife, the
  {B}ootstrap and {O}ther {R}esampling {P}lans}}}\ (\bibinfo  {publisher}
  {Society for Industrial and Applied Mathematics, Philadelphia},\ \bibinfo
  {year} {1982})\BibitemShut {NoStop}%
\bibitem [{\citenamefont {Spirin}\ \emph {et~al.}(2001)\citenamefont {Spirin},
  \citenamefont {Krapivsky},\ and\ \citenamefont {Redner}}]{spirin2001}%
  \BibitemOpen
  \bibfield  {author} {\bibinfo {author} {\bibfnamefont {V.}~\bibnamefont
  {Spirin}}, \bibinfo {author} {\bibfnamefont {P.L.}\ \bibnamefont
  {Krapivsky}}, \ and\ \bibinfo {author} {\bibfnamefont {S.}~\bibnamefont
  {Redner}},\ }\bibfield  {title} {\enquote {\bibinfo {title} {Fate of
  zero-temperature {I}sing ferromagnets},}\ }\href@noop {} {\bibfield
  {journal} {\bibinfo  {journal} {Phys. Rev. E}\ }\textbf {\bibinfo {volume}
  {63}},\ \bibinfo {pages} {036118} (\bibinfo {year} {2001})}\BibitemShut
  {NoStop}%
\bibitem [{\citenamefont {Wu}(1982)}]{Potts_RMP}%
  \BibitemOpen
  \bibfield  {author} {\bibinfo {author} {\bibfnamefont {F.-Y.}\ \bibnamefont
  {Wu}},\ }\bibfield  {title} {\enquote {\bibinfo {title} {The {P}otts
  model},}\ }\href@noop {} {\bibfield  {journal} {\bibinfo  {journal} {Rev.
  Mod. Phys.}\ }\textbf {\bibinfo {volume} {54}},\ \bibinfo {pages} {235}
  (\bibinfo {year} {1982})}\BibitemShut {NoStop}%
\bibitem [{\citenamefont {Majumder}\ \emph {et~al.}(2018)\citenamefont
  {Majumder}, \citenamefont {Das},\ and\ \citenamefont
  {Janke}}]{majumder2018Potts}%
  \BibitemOpen
  \bibfield  {author} {\bibinfo {author} {\bibfnamefont {S.}~\bibnamefont
  {Majumder}}, \bibinfo {author} {\bibfnamefont {S.K.}\ \bibnamefont {Das}}, \
  and\ \bibinfo {author} {\bibfnamefont {W.}~\bibnamefont {Janke}},\ }\bibfield
   {title} {\enquote {\bibinfo {title} {Universal finite-size scaling function
  for kinetics of phase separation in mixtures with varying number of
  components},}\ }\href@noop {} {\bibfield  {journal} {\bibinfo  {journal}
  {Phys. Rev. E}\ }\textbf {\bibinfo {volume} {98}},\ \bibinfo {pages} {042142}
  (\bibinfo {year} {2018})}\BibitemShut {NoStop}%
\bibitem [{\citenamefont {Janke}\ \emph {et~al.}(2021)\citenamefont {Janke},
  \citenamefont {Majumder},\ and\ \citenamefont {Das}}]{Janke_CP}%
  \BibitemOpen
  \bibfield  {author} {\bibinfo {author} {\bibfnamefont {W.}~\bibnamefont
  {Janke}}, \bibinfo {author} {\bibfnamefont {S.}~\bibnamefont {Majumder}}, \
  and\ \bibinfo {author} {\bibfnamefont {S.K.}\ \bibnamefont {Das}},\
  }\bibfield  {title} {\enquote {\bibinfo {title} {Universal finite-size
  scaling function for coarsening in the {P}otts model with conserved
  dynamics},}\ }\href@noop {} {\bibfield  {journal} {\bibinfo  {journal} {J.
  Phys.: Conf. Ser.}\ }\textbf {\bibinfo {volume} {2122}},\ \bibinfo {pages}
  {012009} (\bibinfo {year} {2021})}\BibitemShut {NoStop}%
\bibitem [{\citenamefont {Henkel}\ and\ \citenamefont
  {Pleimling}(2010)}]{henkel_book}%
  \BibitemOpen
  \bibfield  {author} {\bibinfo {author} {\bibfnamefont {M.}~\bibnamefont
  {Henkel}}\ and\ \bibinfo {author} {\bibfnamefont {M.}~\bibnamefont
  {Pleimling}},\ }\href@noop {} {\emph {\bibinfo {title} {Non-{E}quilibrium
  {P}hase {T}ransitions: {A}geing and {D}ynamical {S}caling far from
  {E}quilibrium}}},\ Vol.~\bibinfo {volume} {2}\ (\bibinfo  {publisher}
  {Springer, Heidelberg},\ \bibinfo {year} {2010})\BibitemShut {NoStop}%
\bibitem [{\citenamefont {Midya}\ \emph {et~al.}(2014)\citenamefont {Midya},
  \citenamefont {Majumder},\ and\ \citenamefont {Das}}]{midya2014}%
  \BibitemOpen
  \bibfield  {author} {\bibinfo {author} {\bibfnamefont {J.}~\bibnamefont
  {Midya}}, \bibinfo {author} {\bibfnamefont {S.}~\bibnamefont {Majumder}}, \
  and\ \bibinfo {author} {\bibfnamefont {S.K.}\ \bibnamefont {Das}},\
  }\bibfield  {title} {\enquote {\bibinfo {title} {Aging in ferromagnetic
  ordering: {F}ull decay and finite-size scaling of autocorrelation},}\
  }\href@noop {} {\bibfield  {journal} {\bibinfo  {journal} {J. Phys.: Condens.
  Matt.}\ }\textbf {\bibinfo {volume} {26}},\ \bibinfo {pages} {452202}
  (\bibinfo {year} {2014})}\BibitemShut {NoStop}%
\bibitem [{\citenamefont {Midya}\ \emph {et~al.}(2015)\citenamefont {Midya},
  \citenamefont {Majumder},\ and\ \citenamefont {Das}}]{midya2015}%
  \BibitemOpen
  \bibfield  {author} {\bibinfo {author} {\bibfnamefont {J.}~\bibnamefont
  {Midya}}, \bibinfo {author} {\bibfnamefont {S.}~\bibnamefont {Majumder}}, \
  and\ \bibinfo {author} {\bibfnamefont {S.K.}\ \bibnamefont {Das}},\
  }\bibfield  {title} {\enquote {\bibinfo {title} {Dimensionality dependence of
  aging in kinetics of diffusive phase separation: {B}ehavior of
  order-parameter autocorrelation},}\ }\href@noop {} {\bibfield  {journal}
  {\bibinfo  {journal} {Phys. Rev. E}\ }\textbf {\bibinfo {volume} {92}},\
  \bibinfo {pages} {022124} (\bibinfo {year} {2015})}\BibitemShut {NoStop}%
\bibitem [{\citenamefont {Majumder}\ and\ \citenamefont
  {Das}(2011{\natexlab{b}})}]{majumder2011EPL}%
  \BibitemOpen
  \bibfield  {author} {\bibinfo {author} {\bibfnamefont {Suman}\ \bibnamefont
  {Majumder}}\ and\ \bibinfo {author} {\bibfnamefont {Subir~K}\ \bibnamefont
  {Das}},\ }\bibfield  {title} {\enquote {\bibinfo {title} {Universality in
  fluid domain coarsening: The case of vapor-liquid transition},}\ }\href@noop
  {} {\bibfield  {journal} {\bibinfo  {journal} {Europhysics Letters}\ }\textbf
  {\bibinfo {volume} {95}},\ \bibinfo {pages} {46002} (\bibinfo {year}
  {2011}{\natexlab{b}})}\BibitemShut {NoStop}%
\bibitem [{\citenamefont {Majumder}\ and\ \citenamefont
  {Das}(2013{\natexlab{b}})}]{MajumderPRL}%
  \BibitemOpen
  \bibfield  {author} {\bibinfo {author} {\bibfnamefont {S.}~\bibnamefont
  {Majumder}}\ and\ \bibinfo {author} {\bibfnamefont {S.K.}\ \bibnamefont
  {Das}},\ }\bibfield  {title} {\enquote {\bibinfo {title} {Effects of density
  conservation and hydrodynamics on aging in nonequilibrium processes},}\
  }\href@noop {} {\bibfield  {journal} {\bibinfo  {journal} {Phys. Rev. Lett.}\
  }\textbf {\bibinfo {volume} {111}},\ \bibinfo {pages} {055503} (\bibinfo
  {year} {2013}{\natexlab{b}})}\BibitemShut {NoStop}%
\end{thebibliography}
\end{document}